\documentclass[aps,prb,twocolumn,showpacs,eqsecnum,superscriptaddress,floatfix,10pt]{revtex4-1}

\usepackage{amssymb}  
\usepackage{color,soul} 
\usepackage{graphicx}  
\usepackage[tight]{subfigure}  
\usepackage{mathrsfs}
\usepackage[pdftex,pdfusetitle,bookmarks=true,colorlinks=true,citecolor=blue,urlcolor=blue,linkcolor=magenta]{hyperref}
\usepackage{hypcap}
\usepackage{braket}
\usepackage{xfrac}
\usepackage{tikz}
\usepackage{url}
\usepackage{setspace}
\usepackage{enumerate}
\usepackage{bm}
\usepackage{epstopdf}
\usepackage{amsmath}
\usepackage{amsfonts}
\usepackage{mhchem} 
\usepackage[normalem]{ulem}

\begin{document}

\title{Multiple topological transitions in twisted bilayer graphene near the first magic angle}

\author{Kasra Hejazi}
\thanks{These two authors contributed equally.}
\affiliation{Department of Physics, University of California Santa Barbara, Santa Barbara, California, 93106, USA}
\author{Chunxiao Liu}
\thanks{These two authors contributed equally.}
\affiliation{Department of Physics, University of California Santa Barbara, Santa Barbara, California, 93106, USA}
\author{Hassan Shapourian}
\affiliation{James Franck Institute and Kadanoff Center for Theoretical Physics, University of Chicago, Illinois 60637, USA}
\affiliation{Kavli Institute for Theoretical Physics, University of California Santa Barbara, CA 93106, USA}
\author{Xiao Chen}
\affiliation{Kavli Institute for Theoretical Physics, University of California Santa Barbara, CA 93106, USA}
\author{Leon Balents}
\affiliation{Kavli Institute for Theoretical Physics, University of California Santa Barbara, CA 93106, USA}

\date{\today}
\begin{abstract}
Recent experiments have observed strongly correlated physics in twisted bilayer graphene (TBG) at very small angles, along with nearly flat electron bands at certain fillings. A good starting point in understanding the physics is a continuum model (CM) proposed by Lopes dos Santos \textit{et al.} [Phys. Rev. Lett. 99, 256802 (2007)] and Bistritzer \textit{et al.} [PNAS 108, 12233 (2011)] for TBG at small twist angles, which successfully predicts the bandwidth reduction of the middle two bands of TBG near the first magic angle $\theta_0=1.05^\circ$. In this paper, we analyze the symmetries of the CM and investigate the low energy flat band structure in the entire moir\'e Brillouin zone near $\theta_0$. Instead of observing flat bands at only one ``magic'' angle, we notice that the bands remain almost flat within a small range around $\theta_0$, where \textit{multiple} topological transitions occur. The topological transitions are caused by the creation and annihilation of Dirac points at either $\text{K}$, $\text{K}^\prime$, or $\Gamma$ points, or along the high symmetry lines in the moir\'e Brillouin zone. We trace the evolution of the Dirac points, which are very sensitive to the twist angle, and find that there are several processes transporting Dirac points from $\Gamma$ to $\text{K}$ and $\text{K}^\prime$. At the $\Gamma$ point, the lowest energy levels of the CM are doubly degenerate for some range of twisting angle around $\theta_0$, suggesting that the physics is \textit{not} described by any two band model. Based on this observation, we propose an effective six-band model (up to second order in quasi-momentum) near the $\Gamma$ point with the full symmetries of the CM, which we argue is the minimal model that explains the motion of the Dirac points around $\Gamma$ as the twist angle is varied. By fitting the coefficients from the numerical results, we show that this six-band model captures the important physics over a wide range of angles near the first ``magic'' angle.
\end{abstract}

\maketitle

\section{Introduction}
Recently, the physics of \textit{twisted} bilayer graphene (TBG) has garnered substantial interest both experimentally and theoretically.   For small twist angles a moir\'e pattern with a large unit cell results.   Theoretically, the band structure then becomes very strongly renormalized and sensitive to the precise angle $\theta$ of rotation of one layer relative to the other\cite{2007PhRvL..99y6802L}.  At certain \textit{magic} angles, the Dirac velocity of the bands near charge neutrality is predicted to vanish\cite{Bistritzer2011}, with the entire low energy bands becoming exceptionally flat throughout the moir\'e Brillouin zone.   Experimentally, the renormalization  of electronic structure by the twist has been observed through tunneling measurements\cite{li2010observation}.   It is expected that flat bands promote interaction-induced instabilities, when the band-width becomes comparable or smaller than the interaction energy.  Recently, when the filling is such that  the Fermi level lies within these flat bands, i.e.~close to charge neutrality, correlated insulating states\cite{cao2018correlated,kim2017tunable} and superconductivity \cite{cao2018unconventional} were observed, and similar insulators were observed in trilayers\cite{chen2018gate}. Also new experimental results \cite{2018arXiv180807865Y} have confirmed previously found correlated states in TBG systems. Furthermore, they have observed strong correlations at angles larger than the magic angle range, when sufficient pressure is applied. All these experimental studies have stimulated a large number of theoretical works attempting to explain these phenomena\cite{2018arXiv180308057X,2018PhRvB..98d5103Y,2018arXiv180505294R,2018arXiv180403162D,2018PhRvL.121b6402W,2018arXiv180508232Z,2018arXiv180704382L,2018arXiv180309742P,2018arXiv180507303P,2018arXiv180506867Y,2018arXiv180506906W,2018arXiv180506449I,2018arXiv180410009L,2018arXiv180504918K,2018arXiv180506310K,2018PhRvB..97w5453G,2018JETPL.tmp...82I}.

Here we take a step back and return to the electronic structure of the flat bands.    Theoretically, a continuum description, in which the states are derived primarily from the vicinity of the Dirac points in each layer, is expected to apply when the twist angle is small.  Such a model was derived by Lopes dos Santos {\em et al}\cite{2007PhRvL..99y6802L}, and studied more intensely by Bistritzer and Macdonald\cite{Bistritzer2011}, who discovered the sequence of magic angles.  We will use the Continuum Model (CM) of these authors in this paper.  The CM has an infinite number of bands.  Many authors have sought to simplify further to construct low energy effective models of the flat bands {\em alone}, i.e. containing just two bands (or 4 if the valley degree of freedom is counted).    Such a theory should possess the correct symmetries of TBG \cite{2018arXiv180309742P,2018arXiv180607873Z,2012PhRvB..86o5449L}, and should be able to recover the real space charge density distributions, which is seen to peak at the moir\'e triangular sites \cite{cao2018correlated}. In this regard, Wannier functions for TBG have been studied by several authors \cite{2018arXiv180504918K, 2018arXiv180506819K, 2018arXiv180309742P} as a preliminary step towards a tight binding model for the flat bands. However, the expected symmetry and topology of these flat bands impose obstructions to the use of Wannier functions for a tight binding construction \cite{2018arXiv180607873Z}, to which one remedy is to regard some of the symmetries as emergent ones at low energy, rather than the symmetries of the bilayer system. Due to these complications, finding a low energy non-interacting model describing the bands with correct symmetries is still a subject of intensive research.

The CM still provides a significant simplication compared to a full microscopic treatment.  Specifically, while  a tight binding description has two dimensionless parameters -- the ratio of the inter-layer hopping to the intra-layer one, {\em and} the twist angle -- the CM has only one non-trivial dimensionless parameter $\alpha$, which is given by the ratio of the inter-layer hopping energy to the energy mismatch of the layers due to the shift of the zone boundary wavevector caused by the rotation.  When both of the aforementioned microscopic dimensionless parameters are small, the CM applies, and $\alpha$ can still be arbitrary.  We note that since the inter-layer hopping is essentially fixed physically, $\alpha\sim 1/\theta$. Another significant advantage of the CM is that within it, the angle (and hence $\alpha$) can be tuned continuously, while microscopically, the system is only quasiperiodic in real space for all but a set of measure zero of angles, and hence has no true Brillouin zone.  In the CM the two microscopic Dirac valleys are decoupled, and can be treated independently.  We will focus just on one valley, the other being its time reversal partner.  Within the CM, the Dirac cones of the two layers reconstruct and form (sub-)bands which are perioidic in the moir\'e Brillouin zone (we will use BZ to abbreviate the moir\'e Brillouin zone).  The low energy physics near charge neutrality is governed by the two middle bands, which touch at two effective Dirac cones at the corners of the BZ (we denote these $\text{K}$ and $\text{K}^\prime$ points) for all values of $\alpha$.  The Dirac velocity at these points vanishes for the sequence of magic angles, and moreover, the width of the two middle bands is very small in the vicinity of each of these magic angles, i.e.~for almost all the angles satisfying $ \theta \lessapprox 1^\circ$.   The CM has been the starting point of many works, and the low energy band dispersions in the BZ as a function of the twisting angle $\theta$ have been studied in Ref. \onlinecite{cao2018unconventional}, but the resolution was not fine enough to uncover the entire low energy physics. 

In this paper, we study the low energy band structure of the CM in a more careful manner. 
We find that the low energy flat bands exist not only at the so-called first magic angle for which $\alpha_0=0.6051$, but also within a finite range of $\alpha$ around it.
As we gradually vary $\alpha$ near $\alpha_0$, we observe that (1) the middle two bands remain almost flat and (2) multiple topological transitions occur in the middle two band due to the creation and annihilation of Dirac points (DPs) in the entire BZ. Apart from the two DPs at $\text{K}$ and $\text{K}^\prime$ points which persist at all $\alpha$, DPs also appear at $\Gamma$ point (the center of BZ) for specific $\alpha$ values and along the high symmetry lines in the BZ for a range of $\alpha$ near $\alpha_0$. These DPs carry vorticity (topological charge) and are topologically protected by symmetries of the CM. Furthermore, they can only be annihilated by fusing with the DPs of opposite vorticity. 

We construct an effective model to better understand the low energy physics in the vicinity of the $\Gamma$ point, for $\alpha$ close to $\alpha_0$.  Notice that when $0.57524<\alpha<0.6125$, the middle bands at $\Gamma$ point are each two-fold degenerate, implying that the effective theory should be at least four-dimensional instead of two-dimensional. Away from the above regime, the middle two energy levels are no longer degenerate and carry a different representation of the symmetry group, which suggests that the minimal model for the low energy physics at $\Gamma$ is six-dimensional. We propose such a six-band model using pure symmetry constraints. By including terms up to second order in momentum, this model is capable of capturing the correct number and location of DPs and the rich topological transitions around $\alpha\sim \alpha_0$ predicted by the CM.

The locations and total number of these DPs are highly sensitive to $\alpha$. At some $\alpha<\alpha_0$, twelve DPs emerge around the $\Gamma$ point, and as one increases $\alpha$, six of those with positive topological charge move towards $\text{K}$ and $\text{K}^\prime$ and cause the Lifshitz transition at $\alpha_0$. As the value of $\alpha$ exceeds $\alpha_0$, these twelve DPs continue to move around in the BZ, until $\alpha = 0.74$, at which each one of the positively charged DPs annihilates with one of the negative oness. For $\alpha > 0.74$, one is again left with only the two DPs at $\text{K}$ and $\text{K}^\prime$. The existence of a large number of DPs forces the two bands to repeatedly approach each other at many points in the BZ, which is potentially the ultimate reason for the observation of the nearly flat bands around the first magic angle. 

The rest of the paper is organized as follows. In Sec.~\ref{sec:BM_model}, we introduce the CM and analyze various symmetries satisfied by this model. Then in Sec.~\ref{sec:numerical} we numerically study how the DPs in the middle two bands evolve as one changes $\alpha$ close to the first magic angle. In particular, we discuss the trajectories of the DPs in the BZ, and assign them a vorticity by computing their Berry phase. An effective model for momenta close to $\text{K}$ (and also $\text{K}^\prime$) is also presented, which describes Lifshitz transition at the first magic angle. In Sec.~\ref{sec:six_band}, we introduce the effective six-band model mentioned above to linear and quadratic order in momenta. By numerical fitting, we show that the physics of the emergence of the twelve DPs near $\Gamma$ point exhibited in the CM can be accurately captured by this six band model. Finally, we discuss our results and possible future directions in  Sec.~\ref{sec:conclusion_outlook}.

\section{Continuum model and its symmetries} \label{sec:BM_model}

\begin{figure}[h]
\centering
\subfigure[]{\label{fig:bz_1} \includegraphics[width=.2\textwidth]{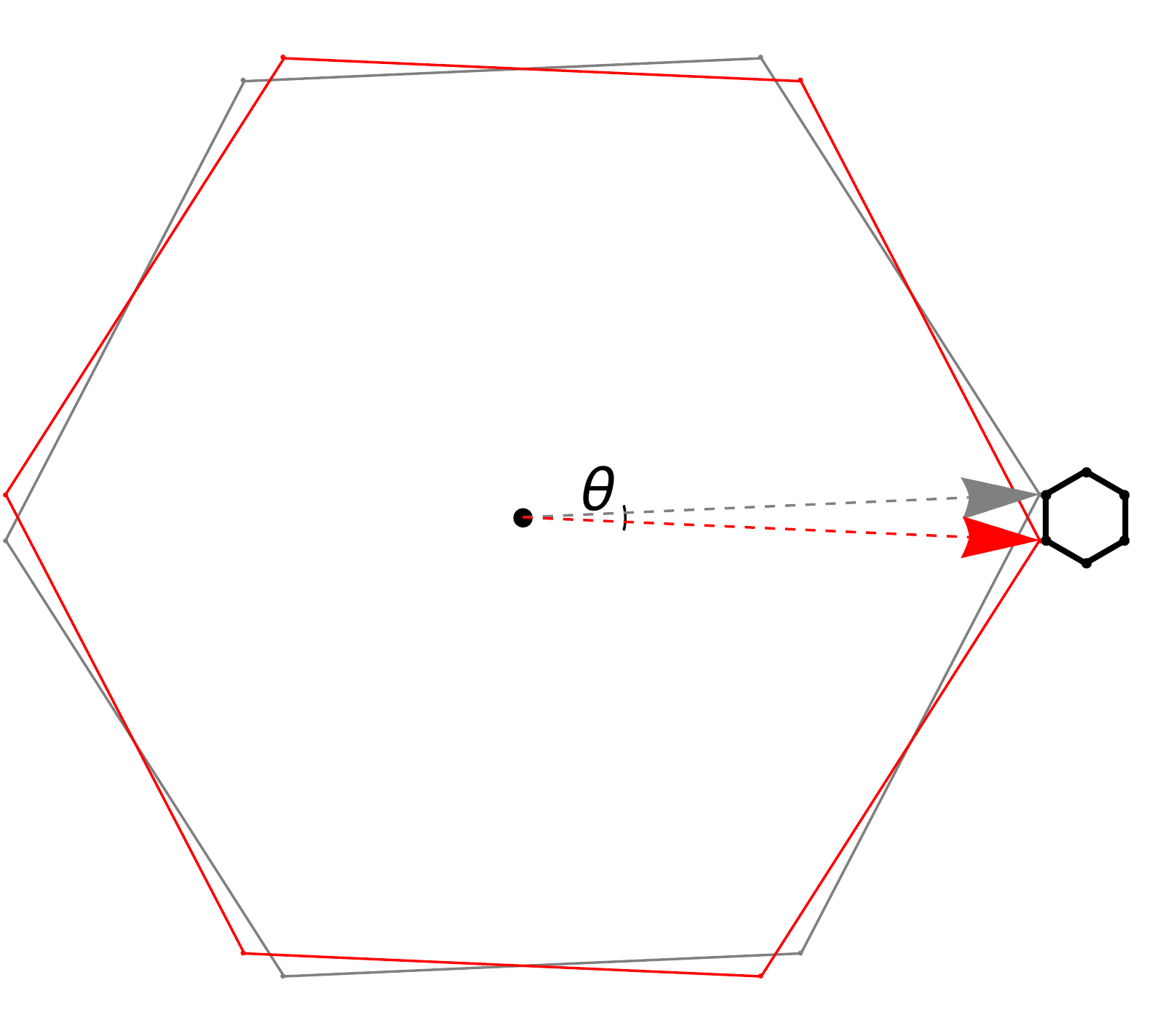}} \quad
\subfigure[]{\label{fig:bz_2} \includegraphics[width=.2\textwidth]{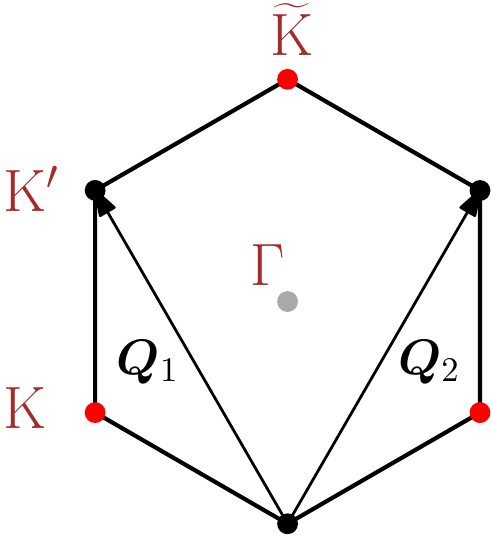}}
\caption{The relevant Brillouin zones. (a) The Brillouin zone for the top (bottom) single layer is shown by the big red (gray) hexagon. The twist angle $\theta > 0$ is labeled; the top layer is rotated by an angle of $-\theta/2$ and the bottom one with $\theta/2$ in real space. The BZ of the TBG system for the valley on the right is shown by the small black hexagon bordering the single layer Brillouin zones; its zoomed-in version is shown in (b), where the high symmetry points $\Gamma$, $\text{K}$ and $\text{K}^\prime$ are labeled. The moir\'e reciprocal lattice vectors $\bm{{Q}}_{1,2}$ are shown. Notice that the $\rm{K}$ point with its other two equivalent points are labeled by red dots and the $\rm{K}^\prime$ point with its equivalent points are labeled by black dots. We further denote ${\bm k}=(0, 1)$ as $\widetilde{K}$ point. In most of this paper, we will focus on the topological transitions near $\Gamma$ and $\widetilde{\rm K}$ points.}
\label{fig:BZ}
\end{figure}

We use a slight reformulation of the CM as presented by Bistritzer and Macdonald in Ref.~\onlinecite{Bistritzer2011}.  In their work, a separate momentum space origin was chosen for each layer (to coincide with the layer's DP).  We prefer to use a common origin, so momentum has the conventional meaning (see Fig.~\ref{fig:bz_1}).  A layer-dependent unitary rotation of their Hamiltonian, written in real space, accomplishes this, and produces the form convenient for our further analysis:
\begin{eqnarray}
\label{eq:TBGa}
H(\bm{x}) = &&-i\left(\bm{\nabla} - i\tau^z \frac{\bm{q}_0}{2} + i\bm{q}_{\text{h}}\right)\cdot\bm{\sigma}
\nonumber\\
&&+ \alpha \; \tau^+ \left[\alpha(\bm{x})+\beta(\bm{x}) \sigma^+ + \gamma(\bm{x}) \sigma^-\right] +\mathrm{h.c.},\nonumber\\
\end{eqnarray}
where $\bm{q}_{\text{h}}=\left(\frac{\sqrt{3}}{2},0\right)$, and $\bm{q}_0 = \left( 0 , -1 \right)$ represents the shift of one DP relative to another by the rotation, in units scaled by the size of the rotation (see below).  We use Pauli matrix notations to address the layers -- $\tau^z=\pm 1$ for the top and bottom layers -- and the sublattice $\sigma^z=\pm 1$ for the A and B sublattices.  The tunneling part of the Hamiltonian (i.e. the $\tau^+ = \frac{1}{2}(\tau^x + i \tau^y)$ term and its hermitian conjugate) contains three functions
\begin{subequations}\label{eq:abc}
\begin{eqnarray}
\alpha(\bm{x}) &=& \sum\limits_{j=0}^2 e^{-i\bm{{Q}}_j\cdot \bm{x}},\\
\beta(\bm{x}) &=& \sum\limits_{j=0}^2 e^{-i\bm{{Q}}_j\cdot \bm{x}}\zeta^j,\\
\gamma(\bm{x}) &=& \sum\limits_{j=0}^2 e^{-i\bm{{Q}}_j\cdot \bm{x}}\bar{\zeta}^j.
\end{eqnarray}
\end{subequations}
Here $\zeta = e^{2\pi i/3}$, $\bm{{Q}}_{0} = \left(0,0\right)$, $\bm{Q}_1 = \sqrt{3} \left(-\frac{1}{2},\frac{\sqrt{3}}{2}\right)$ and $\bm{Q}_2 = \sqrt{3} \left(\frac{1}{2},\frac{\sqrt{3}}{2}\right)$ are the reciprocal lattice vectors for the BZ (see Fig.~\ref{fig:bz_2}). Note that we have conducted a real space rescaling $\bm{x}\rightarrow \bm{x}/ k_\theta$ so that units are chosen proportionally to the size of the moir\'e unit cell.  Energy is measured in units of $v k_\theta$, which absorbs the Dirac velocity, $v$.  The angle dependence is now subsumed into the parameter $\alpha = \frac{w}{ v k_\theta}$, where $k_\theta = \frac{4\pi}{3\sqrt{3} a} \theta $ ($\theta$ is assumed small) is the displacement between the $\text{K}$ points of the Brillouin zones of the two layers, $a$ is the spacing between two nearest carbon atoms in a monolayer graphene, and $w$ is the tunneling strength in Bistritzer and MacDonald's original notation. Note that while $\theta$ is the only physically tunable parameter, the rescaled model~\eqref{eq:TBGa} depends on $\theta$ only through the explicit dependence on $\alpha$, which is inversely proportional to $\theta$.  
We remark that in the Dirac term in~\eqref{eq:TBGa} the sublattice matrices $\bm{\sigma}$ in principle should be rotated by $-\theta/2$ for the top layer and $+\theta/2$ for the bottom layer, in response to the real space graphene layer twist (see Appendix~\ref{sec:geometry}). However, this angular dependence in the Dirac term was shown to be negligible in Ref.~\onlinecite{Bistritzer2011}, as $\theta$ is a small number of order $\sim 1^\circ\sim 0.02$; since the effect of this rotation is perturbative, we will neglect it in this paper. For a discussion of the effects of including this correction see Sec.~\ref{sec:conclusion_outlook}.

For a complete low energy description of a TBG system one needs to consider the physics given close to both of the two distinct valleys; however, as we mentioned earlier, in the small angle limit the coupling between the two valleys is negligible. Thus one needs to consider two decoupled copies of the Hamiltonian given in \eqref{eq:TBGa} (one of them with slight modifications, see Appendix~\ref{sec:geometry}). As a result of all this, we will focus on one of the valleys throughout this work. Note that within our settings the energy bands for the other valley will be given by rigid transformations of the present ones.

The CM is manifestly periodic, and so we can apply Bloch's theorem.  Writing $H_{\bm{k}} = e^{-i\bm{k}\cdot\bm{x}} H e^{i\bm{k}\cdot\bm{x}}$, where $\bm{k}$ is the quasi-momentum defined in the BZ, we have
  \begin{eqnarray}
    \label{eq:TBG_k}
H_{\bm{k}}(\bm{x}) = &&-i\left(\bm{\nabla} - i\tau^z \frac{\bm{q}_0}{2} + i\bm{q}_{\text{h}}+i\bm{k}\right)\cdot\bm{\sigma}\nonumber\\
&&+ \alpha \; \tau^+ \left(\alpha(\bm{x})+\beta(\bm{x}) \sigma^+ + \gamma(\bm{x}) \sigma^-\right) +\mathrm{h.c.}.\nonumber\\
  \end{eqnarray}
The CM is simple in the sense that it is a non-interacting model with only one tuning parameter, $\alpha$, however the physics it exhibits is far from being understood. Of particular interest is the flat band physics at specific values of $\alpha$: at $\alpha = 0.605, \, 1.2, \cdots$, where the Fermi velocity is renormalized to zero. A first step in understanding this is a symmetry study of the CM.

This model has five explicit symmetries:

\begin{itemize}
\item Translation symmetry in real space:
\begin{equation}
H_{\bm{k}}(\bm{x}+\bm{r}_i) = H_{\bm{k}}(\bm{x}),\quad i=1,2,
\end{equation}
where $\bm{r}_1 = \frac{4\pi}{3} \left( -\frac{\sqrt{3}}{2} , \frac{1}{2} \right), \bm{r}_2 = \frac{4\pi}{3} \left( \frac{\sqrt{3}}{2} , \frac{1}{2} \right)$ are the moir\'e lattice vectors. Dual to this, we have a $\bm{k}$ space translation symmetry along $\bm{Q}_1$, $\bm{Q}_2$ directions: 
\begin{equation}
U^\dag_{T_i}(\bm{x})H_{\bm{k}+\bm{Q}_i}(\bm{x}) U^{\phantom\dag}_{T_i}(\bm{x}) = H_{\bm{k}} (\bm{x}),\quad i=1,2,
\end{equation}
where $U_{T_i} = e^{-i \bm{Q}_i\cdot \bm{x}}$. Unlike the real space translation symmetry, the momentum space translation is not a physical symmetry since it is a direct consequence of the Bloch theorem. Note that although formally we have $e^{i \bm{q}\cdot\bm{x}}H_{\bm{k}+\bm{q}}(\bm{x}) e^{-i\bm{q}\cdot \bm{x}} = H_{\bm{k}} (\bm{x})$ for any momentum $\bm{q}$, the fact that $H_{\bm{k}}(\bm{x})$ is a Bloch Hamiltonian only allows a periodic $U_{T_i}(\bm{x})$ in real space, i.e. $\bm{q} = n_1 \bm{Q}_1 + n_2 \bm{Q}_2$. 

\item $C_3$ rotation symmetry around origin (an $AA$ point, see Appendix~\ref{sec:geometry} for the geometry of TBG) in real space:
\begin{equation}
 U^\dag_{C_3}(\bm{x}) H_{C_3(\bm{k})}(C_3(\bm{x})) U^{\phantom\dag}_{C_3}(\bm{x}) = H_{\bm{k}}(\bm{x}),
\end{equation}
where $C_3$ is the counter-clockwise $2\pi/3$ rotation, and $U_{C_3}(\bm{x}) = e^{i\frac{2\pi}{3} \sigma^z} e^{i\bm{Q}_1\cdot\bm{x} P_+}e^{i2\bm{q}_{\text{h}}\cdot \bm{x}}$, where $P_+ = \frac{1 + \tau^z}{2}$. Note that $C_3$ rotates real space $\bm{x}$ and quasi-momentum $\bm{k}$ in the same way. The $\bm{k}$ translation symmetry and $C_3$ rotation together allow for a new $C'_3$ rotation symmetry around the rotation center $\text{K}$ and $\text{K}^\prime$, a fact that will be used in the $\text{K}$ point representation analysis.

\item Mirror symmetry along the $x=0$ line $M_y\colon (x,y)\rightarrow (x,-y)$, $\left( k_x , k_y \right)  \rightarrow \left( k_x, - k_y \right)$:
\begin{equation}
U^\dag_{M_y} H_{M_y(\bm{k})}(M_y(\bm{x})) U^{\phantom\dag}_{M_y} = H_{\bm{k}}(\bm{x}),
\end{equation}
where $U_{M_y} = \sigma^x\tau^x$.
\item Composition of a $C_2$ rotation and time reversal $\mathcal{T}$:
\begin{equation}
U^\dag_{C_2\mathcal{T}} H^*_{\bm{k}}(-\bm{x})U^{\phantom\dag}_{C_2\mathcal{T}} = H_{\bm{k}}(\bm{x}),
\end{equation}
where $U_{C_2\mathcal{T}} = \sigma^x$ and the $C_2\mathcal{T}$ symmetry is defined by $U_{C_2\mathcal{T}}\mathcal{K}$. 
Note that the full TBG system has both $C_2$ rotation symmetry and time reversal symmetry $\mathcal{T}$ separately. However, each of these symmetry transformations maps the two valleys into each other (see Appendix \ref{sec:geometry}). Since we are focusing on one of the valleys here, we need to consider the composition of the two transformations to stay within the space of one valley.

\item Particle-hole symmetry $\mathcal{C}$:
\begin{equation}
U^\dag_{\text{ph}} H_{(-k_x,k_y)}((-x,y))U^{\phantom\dag}_{\text{ph}} = -H_{\bm{k}}(\bm{x}),
\end{equation}
where $U_{\text{ph}} = \sigma^x \tau^z e^{2i \bm{q}_{\text{h}}\cdot\bm{x}}$. 
The particle-hole symmetry emerges only in the small $\theta$ approximation: it is an artifact of the CM in which the $\theta$ rotation for the sublattice matrices in the Dirac term is neglected. Since the numerical value of $\theta$ is very small ($\sim 0.02$), the particle-hole symmetry introduced above remains a good approximate symmetry. If on the other hand this $\theta$ rotation is retained in the Hamiltonian, it introduces a small asymmetry between the particle and hole bands, and slightly modifies the low energy topological transitions (see the subsequent sections and especially Sec. \ref{sec:conclusion_outlook}).

\end{itemize}
All the four point-group symmetries introduced here act on-site at the $\Gamma$ point in the $\bm{k}$-space (see figure \ref{fig:bz_2}), making this point have the highest symmetry. Before analyzing the rich physics exhibited at this point, let us first look at $\text{K}$ point (and the physics at $\text{K}^\prime$ is ensured to be the same by the mirror symmetry $M_y$). 

Three out of the four symmetries map $\text{K}$ to $\text{K}$: $C'_3$, $C_2\mathcal{T}$ and particle-hole symmetry $\mathcal{C}$. The most prominent feature of $\text{K}$ point is that it remains gapless for all values of $\alpha$ with a Dirac-like dispersion at its vicinity. While there is no algebraic proof for the persistence of zero energy states, a topological and group-theoretical argument has been given in Ref. \onlinecite{2018arXiv180309742P}, and a low energy effective model has been proposed to understand the Dirac-like dispersion and the ``trigonal warping.'' \cite{cao2018correlated}

We can use the symmetry and representation analysis to better understand the degeneracies. Since all the irreducible representations (irreps) are one-dimensional, at $\text{K}$ point we can diagonalize the action of $C_3$ to get an eigenstate of $C_3$: call it $\Psi$. Under $C_3$ rotation, we have
\begin{equation}
\Psi(\bm{x}) \rightarrow U_{\text{K},C_3}(\bm{x}) \Psi(C_3(\bm{x})) = e^{i\varphi} \Psi(\bm{x}),
\end{equation}
where $\varphi$ is some global phase. Noting that $(C_3)^3 = 1$ we must have $e^{3i \varphi} = 1$, $\varphi = \frac{2\pi n}{3}$, $n = 0,\pm 1$. Then, the state $C_2\mathcal{T}  \Psi$ transforms under $C_3$ as
\begin{equation}
\begin{aligned}
	U_{C_2\mathcal{T}}\Psi^*(-\bm{x}) \rightarrow \;
	& U_{\text{K},C_3}(\bm{x})U_{C_2\mathcal{T}} \Psi^*\left( -C_3(\bm{x})\right) \\
	= & U^{\phantom{*}}_{C_2\mathcal{T}} \; U^*_{\text{K},C_3}(-\bm{x}) \Psi^*\left(C_3(-\bm{x})\right) \\
 =&  e^{-i \varphi} \; U_{C_2\mathcal{T}}\Psi^*(-\bm{x}).
\end{aligned}
\end{equation}
This shows that $\Psi' = U_{C_2\mathcal{T}}\Psi^*(-\bm{x})$ is an eigenstate of $C_3$ with opposte $\varphi$. Furthermore, if one zero mode has $C_3$ rotation value $n=1$, then $C_2\mathcal{T}$ symmetry ensures the existence of another zero mode with $n=-1$. The two states form a two-dimensional irrep of the group $\langle C_3,C_2\mathcal{T}\rangle\cong D_3$. The two-dimensional irrep has indeed been confirmed in numerics, see next section.

\section{Numerical solution}\label{sec:numerical}
\subsection{The Lifshitz transition at $\text{K}$ and $\text{K}^\prime$ points}

\begin{figure}[!t]
\centering
 \subfigure[]{\label{fig:alpha_06} \includegraphics[width=.23\textwidth]{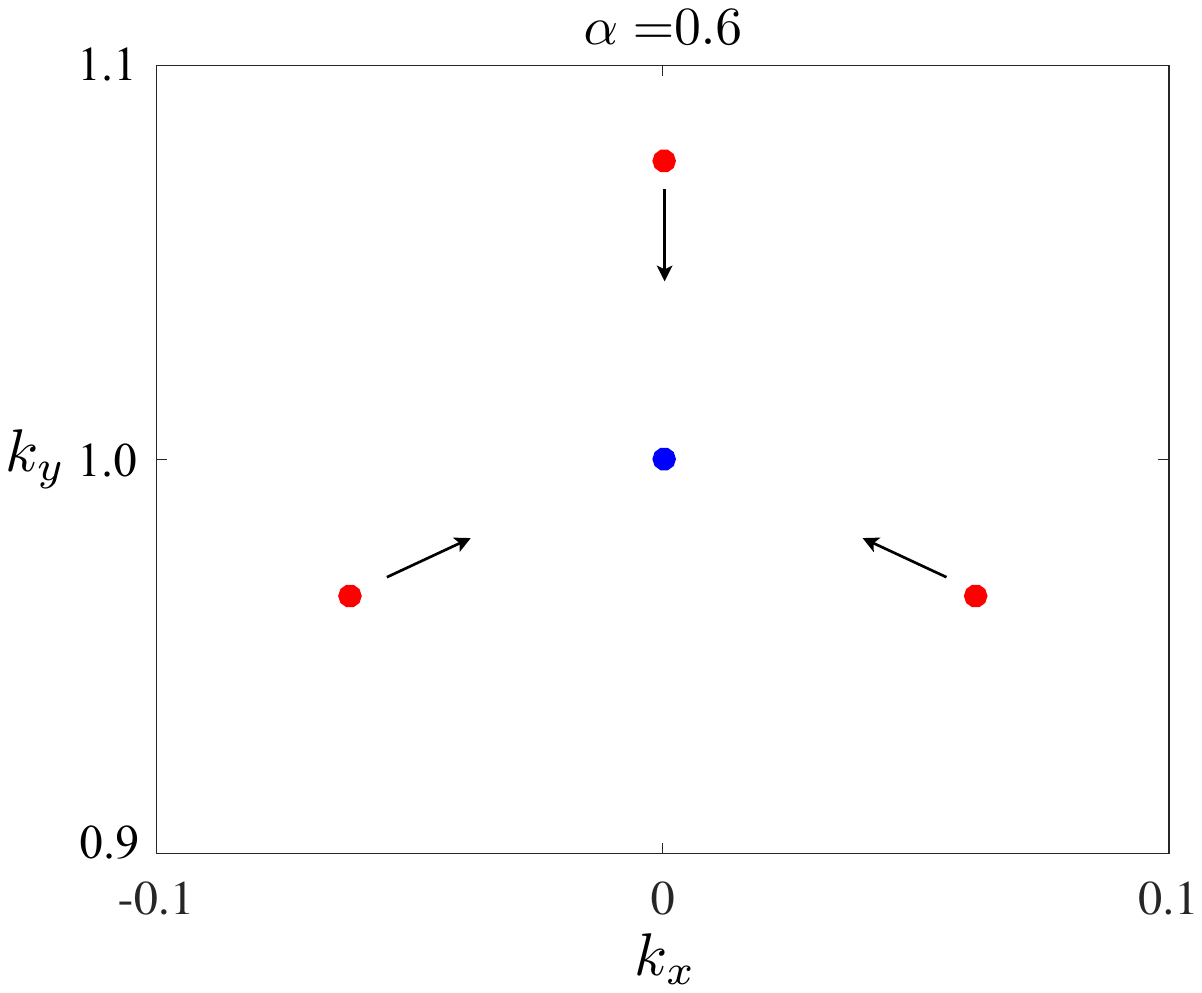}} 
 \subfigure[]{\label{fig:alpha_061} \includegraphics[width=.23\textwidth]{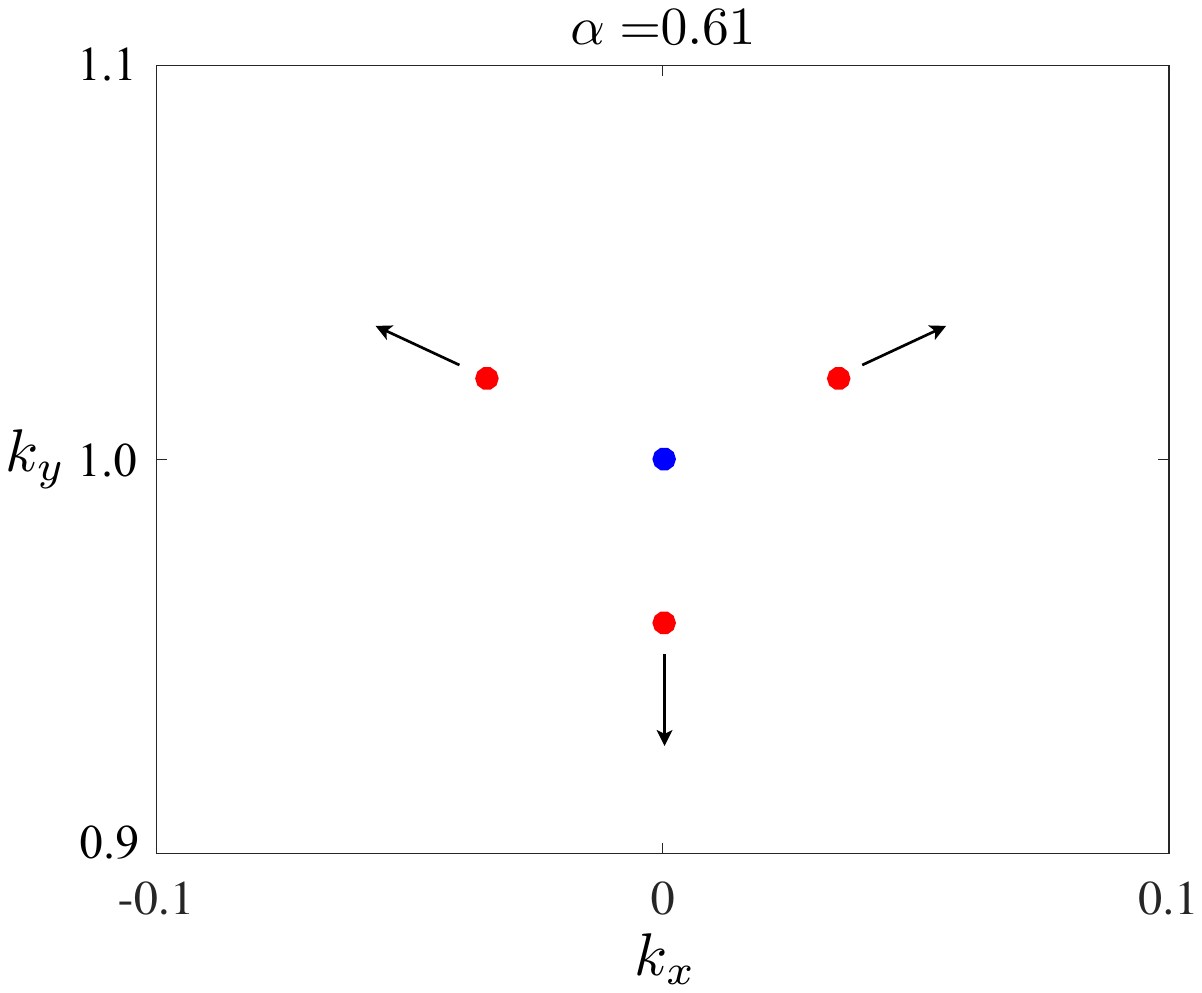}}
 \subfigure[]{\label{fig:Band_alpha_0_3D} \includegraphics[width=.23\textwidth]{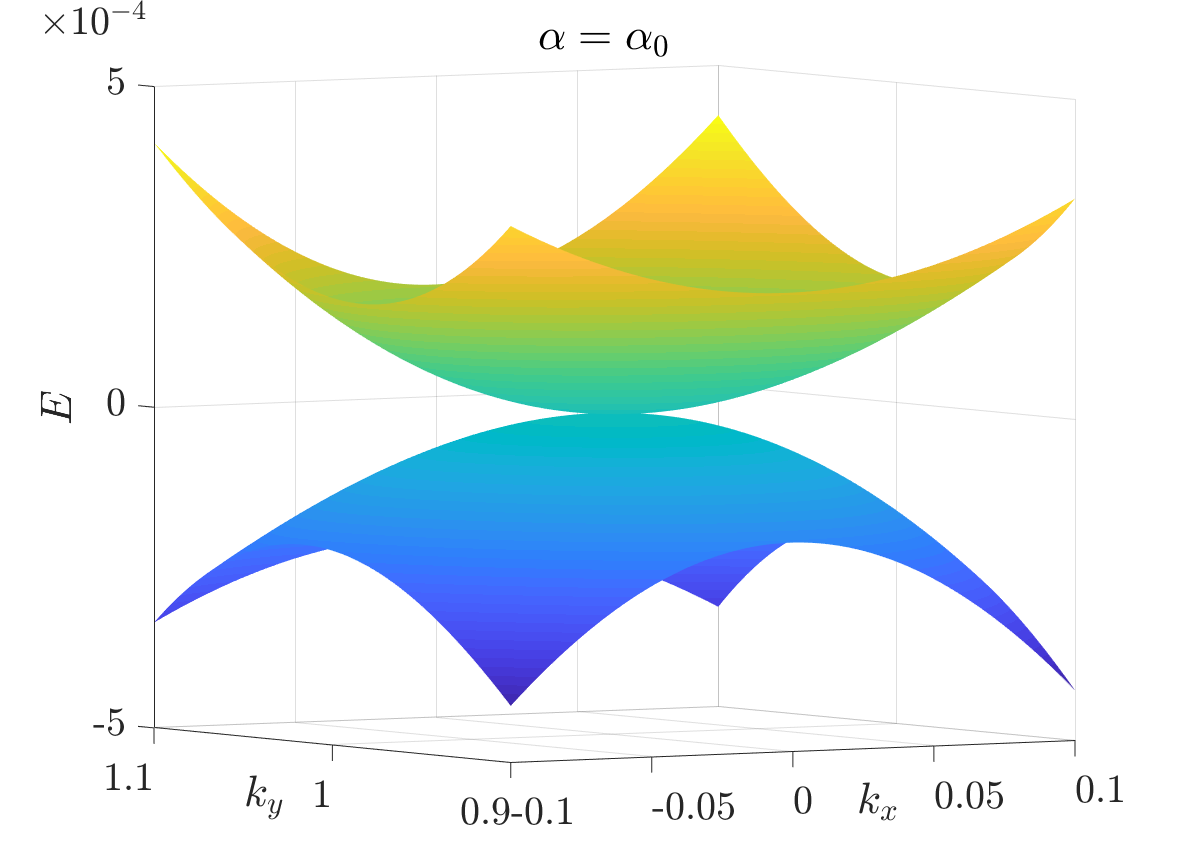}}
 \subfigure[]{\label{fig:Band_alpha_0} \includegraphics[width=.23\textwidth]{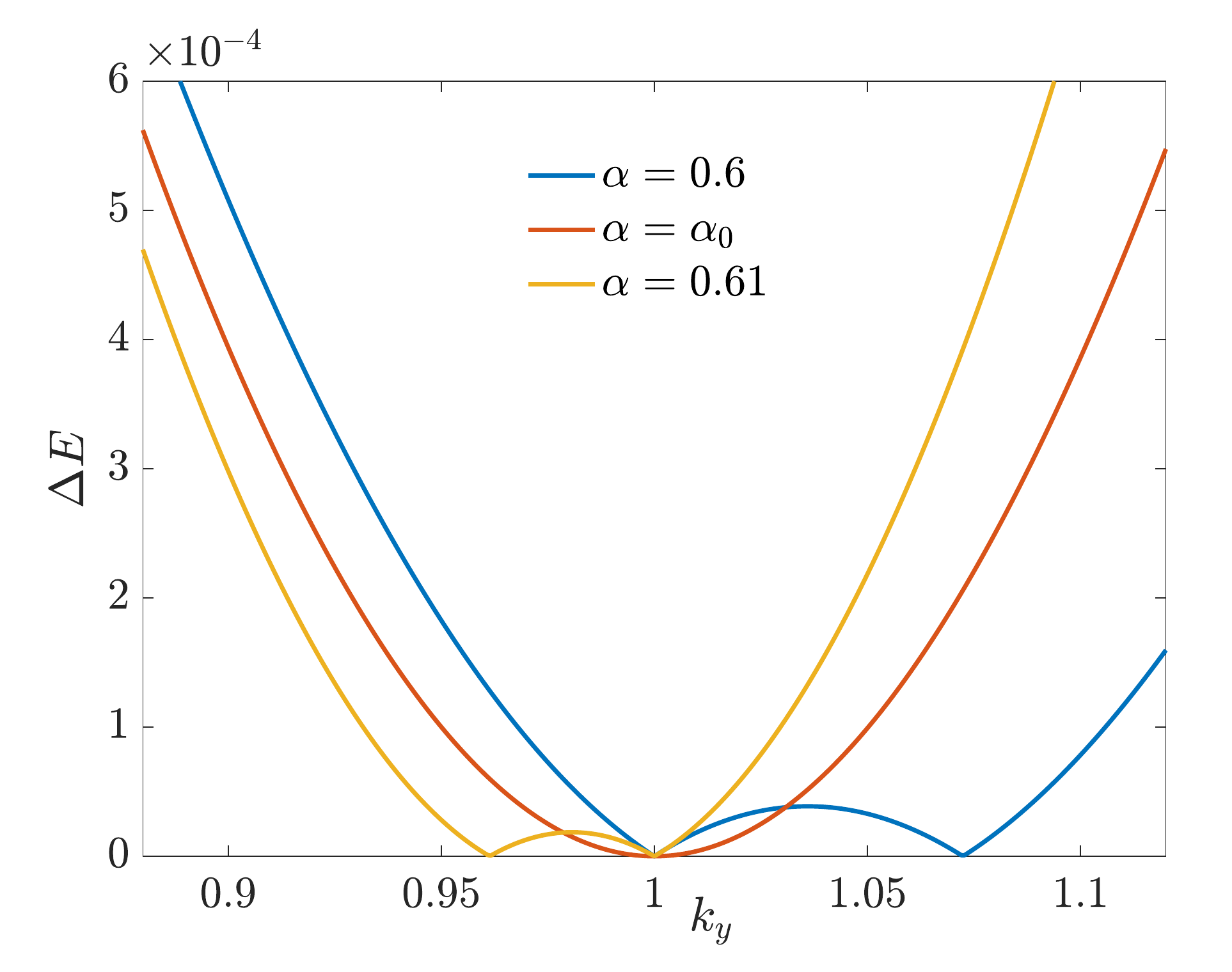}}
\caption{  The Lifshitz transition in the vicinity of  the first magic angle $\alpha_0$. (a) At $\alpha=0.6$, there is one DP with negative topological charge at $\widetilde{\text{K}}$ point (the blue dot, which is equivalent to the $\text{K}$ point) surrounded by three DPs with positive topological charge (red dots). (b) Location and topological charge of DPs at $\alpha=0.61$. 
(c) The quadratic band touching at $\widetilde{\text{K}}$ point for $\alpha=\alpha_0 = 0.6051$. (d) The energy difference between lowest two bands along $k_y$ direction for various $\alpha$ at $k_x=0$. In panels (a) and (b), arrows show the motion of DPs as $\alpha$ is increased. The DPs meet at $\widetilde{\rm K}$ point when $\alpha=\alpha_0$.} 
\label{fig:alpha_0}
\end{figure}

In this subsection, we investigate the middle two bands (flat bands close to $E=0$) around $\text{K}$ or $\text{K}^\prime$ point (and the other equivalent points shown in Fig.~\ref{fig:bz_2}). As discussed in the previous section, at these points, the middle two eigenvalues are two-fold degenerate and are always equal to zero due to the particle-hole symmetry. We focus on the $\widetilde{\text{K}}$ point here, defined in figure \ref{fig:bz_2}, for the ease of notation. At the first magic angle $\alpha_0=0.6051$, near $\widetilde{\text{K}}$, the spectrum has a quadratic dispersion and can be described by a quadratic band touching (QBT) model with Berry phase $2\pi$,
\begin{align}
H_{\rm QBT}\propto \begin{pmatrix} 0 & k_+^2\\
k_-^2 & 0\end{pmatrix},
\end{align}
where $k_{\pm}=k_x\pm ik_y$, and $\bm{k}$ shows the quasi-momentum deviation form the $\widetilde{\text{K}}$ point. Away from $\alpha_0$, this quadratic band touching point breaks into four DPs. This is the so-called trigonal warping (See Fig.~\ref{fig:alpha_0}) in which one DP with negative topological charge sitting at $\widetilde{\text{K}}$ point is surrounded by another three DPs with positive topological charge (See Appendix~\ref{sec:berry_phase} for the discussion of topological charge). The effective Hamiltonian can be considered as $H=H_{\rm QBT}+H_{\rm Dirac}$, i.e., 
\begin{align}
H=  \left( v k_ x + 2 a k_x k_y \right) \mu^1 + \left( v k_y + a \left[ k_x^2 - k_y^2 \right] \right)\mu^2 ,
\end{align}
where $\mu^{1,2}$ are Pauli matrices. This Hamiltonian satisfies all the three symmetries of $\widetilde{\text{K}}$, i.e.~$C_3$, $C_2\mathcal{T}$ and $\mathcal{C}$, discussed in the previous section. Note that $a$ is the curvature of the quadratic band and $v/a$ determines the distance of the three DPs from $\widetilde{\text{K}}$ point. The Lifshitz transition occurs at $v=0$, at which the Hamiltonian becomes $H_{\rm QBT}$. 

Having understood the physics around $\text{K}$ and $\text{K}^\prime$ points, we may ask one question: are there other DPs in the BZ? Since the total topological charge should be conserved, the appearance of three DPs around $\text{K}$ and $\text{K}^\prime$ points suggests that there should be other DPs in the BZ if these two bands are separated from other bands. This motivates us to study the moir\'e band structure of the entire BZ, which is the subject of the next subsection. 

\subsection{The numerics of the Dirac points}
\begin{figure*}
\centering
\includegraphics[width=.75\textwidth]{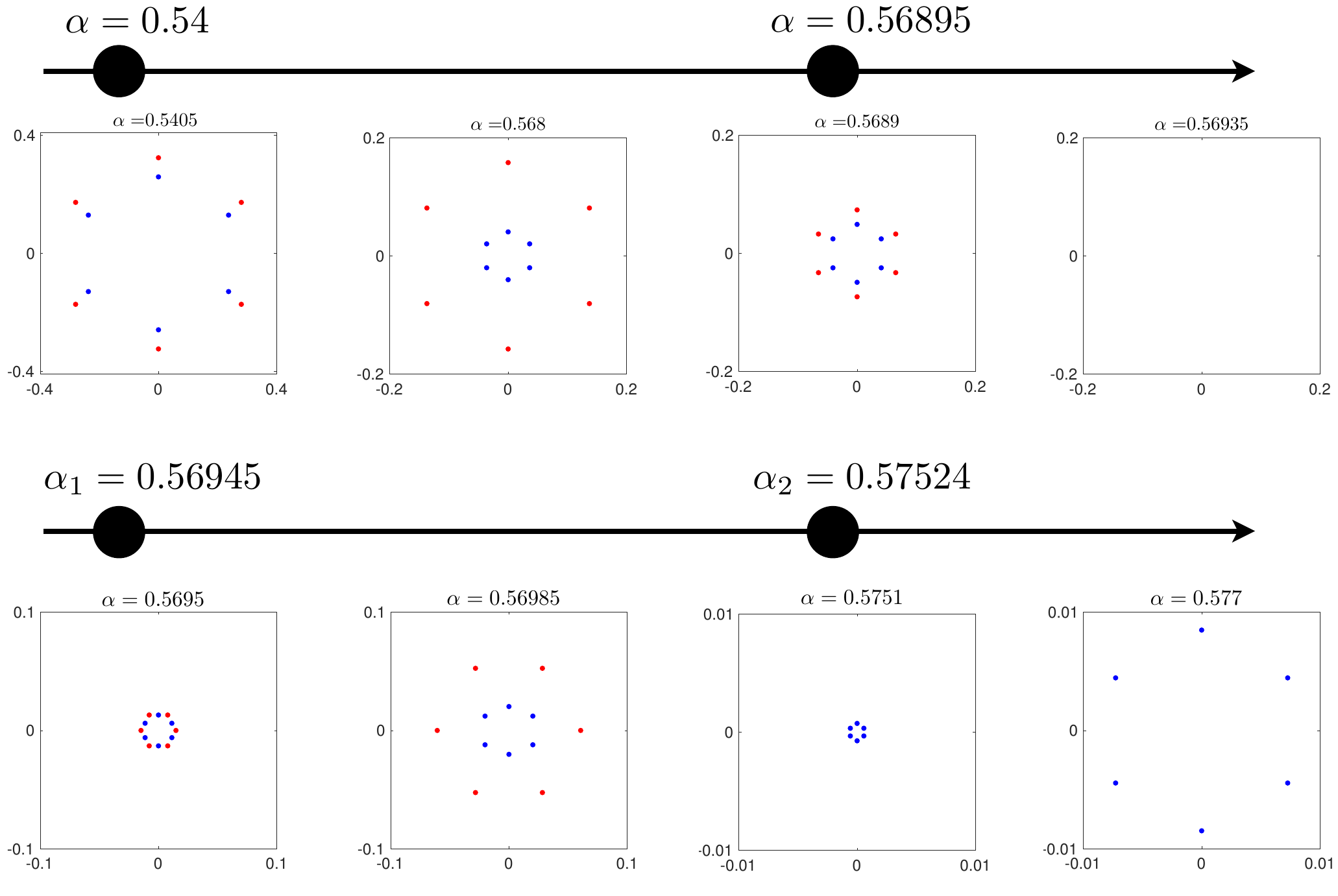}
\caption{The evolution of the DPs around $\Gamma$ point (the center of the eight plots) as we vary $\alpha$. The red dot is the DP with positive charge ($+1/2$ vorticitiy) while the blue dot is the DP with negative charge ($-1/2$ vorticitiy).}
\label{fig:schematic_R}
\end{figure*}

In this subsection, we are going to explore the structure of the DPs of the middle two bands in the entire BZ in the vicinity of $\alpha_0$.

We present the numerical results in Fig.~\ref{fig:schematic_R}. When $\alpha<0.54$, there are only two DPs at $\text{K}$ and $\text{K}^\prime$ points. At $\alpha=0.54$, We notice that six band touching points occur around $\Gamma$ point (the center of BZ) with quadratic dispersion (See Fig.~\ref{fig:schematic_R}). As we increase $\alpha$, each of them splits into two DPs with opposite topological charges. These two DPs annihilate again at $\alpha=0.56895$ and for $\alpha>0.56895$, a small gap forms near $\Gamma$ point.

\begin{figure}[!b]
\centering
\includegraphics[width=.3\textwidth]{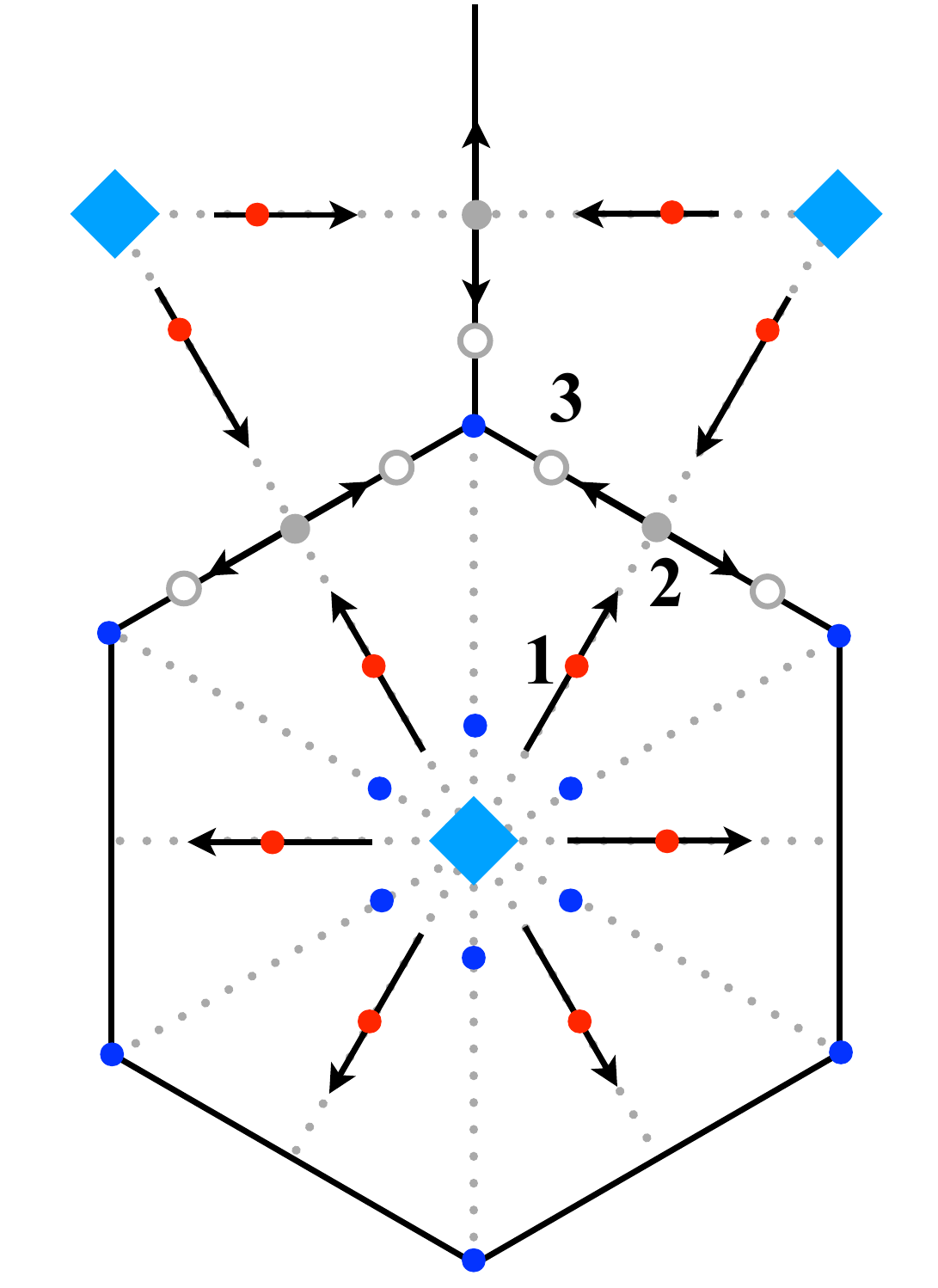}
\caption{The schematics of DPs around $\Gamma$ point. The blue diamonds denote the $\Gamma$ point and two other equivalent points. The red dots represent the DPs with positive charge and blue dots are the DPs with negative charge. The six blue dots are always at zero energy, while the six red dots are not  necessarily at zero energy since none of them are along $k_x=0$ axis (or $C_3$ equivalent directions). However, as they move towards the boundary (gray dots) and further move along the boundary (open circles), they approach zero energy. The arrows denote the motion of DPs as we increase $\alpha$.}
\label{fig:Dirac_motion}
\end{figure}

This small gap closes at $\alpha=0.56945$, where the two bands touch at $\Gamma$ point with quadratic dispersion. Upon further increasing $\alpha$, we find that there are twelve DPs with alternating topological charge at angles (with respect to $k_x$ axis) $2\pi n/12$ ($n=0,1, \dots 11$)  surrounding $\Gamma$ point.  
As $\alpha$ increases, the six DPs with positive charge at even $n$ move away from $\Gamma$ point. Notice that these six DPs are not pinned to zero energy. Three DPs with $n=0,4,8$ (related with $C_3$ symmetry) are at $E=\epsilon>0$ and the other three are particle-hole pairs of them at $E=-\epsilon<0$. If we increase $\alpha$, these six DPs move towards the $\text{K}$ and $\text{K}^\prime$ points. In Fig.~\ref{fig:Dirac_motion}, we explicitly demonstrate the trajectory of the DP at angle $\pi/3$. It involves three steps: 
\begin{enumerate}
\item the DP moves along the $\pi/3$ direction towards the boundary of the BZ.
\item The DP combines with another DP from the neighboring BZ at the boundary (M point of the BZ) and they fuse into a QBT. 
\item This QBT splits into two DPs moving in the opposite directions along the boundary. Eventually, they move towards $\text{K}$ and $\text{K}^\prime$ points and cause the Lifshitz transition over there at $\alpha_0$. 
\end{enumerate}

The other six DPs with negative charge remain close to $\Gamma$ point and are pinned to zero energy due to particle-hole symmetry. Two of them are located along $\pm k_y$ direction with $k_x=0$ and the other four are $C_3$ rotation counterparts. At $\alpha=0.57524$, these six DPs combine at the  $\Gamma$ point, forming a very sharp band touching point, involving the middle four bands. As we further increase $\alpha$, we again find six DPs surrounding $\Gamma$ point along the same directions. At $\alpha=0.74$, these six DPs eventually annihilate with the other six DPs at six points located on the high symmetry lines connecting $\Gamma$ point to $\text{K}$, $\text{K}^\prime$ and the other equivalent  points.

\begin{figure}
\centering
\includegraphics[width=.4\textwidth]{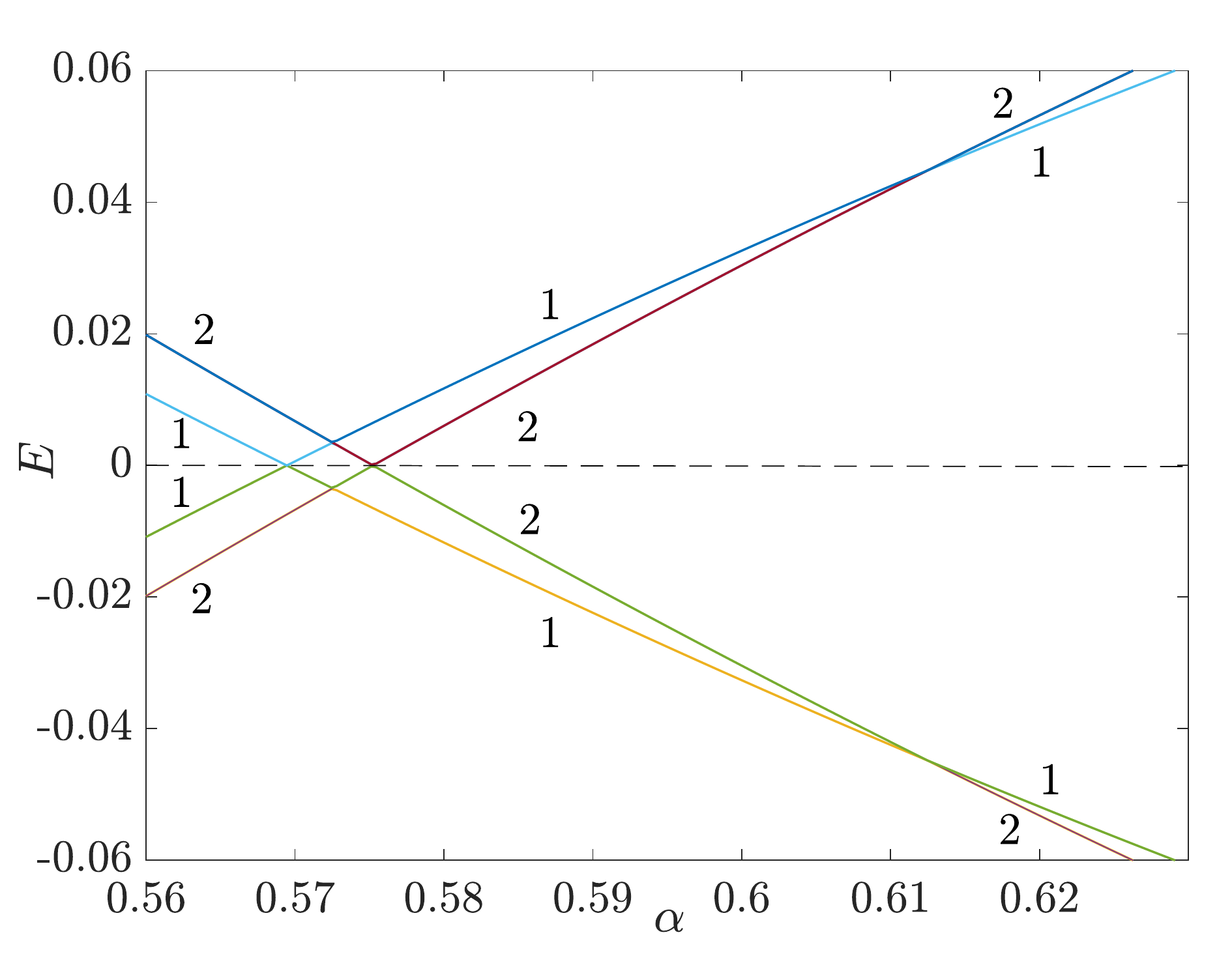}
\caption{The six energy levels closest to $E=0$ at $\Gamma$ point as a function of $\alpha$. The numbers $1$ and $2$ denote the degeneracy. When $\alpha<\alpha_1=0.56945$ and $\alpha>0.6125$, the levels closest to zero are non-degenerate and the higher energy levels are two-fold degenerate. The zero crossing at $\alpha_1 = 0.56945$ is a crossing of non-degenerate levels, while the crossing at $\alpha_2=0.57524$ is a crossing of two-fold degenerate levels.}
\label{fig:Gamma_alpha}
\end{figure}

To better understand the physics around the $\Gamma$ point, we plot the energy levels at $\Gamma$ point as a function of $\alpha$ in Fig.~\ref{fig:Gamma_alpha}. We find that the energy is only equal to zero at two $\alpha$'s and denote them as $\alpha_1=0.56945$ and $\alpha_2=0.57524$, which is also  consistent with the result presented in Fig.~\ref{fig:schematic_R}. The zero crossing at $\alpha_1$ is a crossing of non-degenerate levels, while the crossing at $\alpha_2=0.57254$ is a crossing of two-fold degenerate levels. Moreover, we find that when $\alpha\in[0.5, 0.75]$, the lowest six levels are well separated from the higher levels (with an energy gap of $O(1)$). This motivates us to construct an effective six band Hamiltonian to explain the rich physics around the $\Gamma$ point. 

Before moving on to the next section, let us mention that six points of accidental crossings of the two middle bands also appear when $\alpha$ is very close to $\alpha_2$ on both sides of it. These six nodes are along $\pm k_x$ directions and the other $C_3$ equivalent directions and one finds out that they are topologically trivial if one computes the Berry phase associated with them. As $\alpha$ gets very close to $\alpha_2$ from either higher or lower values, one observes that these crossing points also get very close to the $\Gamma$ point. Please note that these accidental nodes are not shown in Fig. \ref{fig:schematic_R}.

\section{Six band model}\label{sec:six_band}

In this section, we construct a six-band model which describes the low energy behavior around $\Gamma$ point over the parameter range $\alpha\in(0.568, 0.578)$,  including both values $\alpha_1$ and $\alpha_2$.

\subsection{Symmetry and representation analysis}

The $\Gamma$ point is invariant under all the four symmetry operators: $C_3$, $M_y$, $C_2\mathcal{T}$ and $\mathcal{C}$. The six lowest bands form a representation of the group generated by these symmetries; we use this fact to construct an effective Hamiltonian. 

The $C_3$ and $M_y$ symmetry operators generate a six-dimensional group isomorphic to the dihedral group $D_3$, and hence have both one- and two-dimensional irreps. The particle-hole symmetry $\mathcal{C}$ further maps a $+\varepsilon$ state to a $-\varepsilon$ state, doubling the representation dimension. This is consistent with the degeneracies seen in the spectrum in Fig.~\ref{fig:Gamma_alpha}, suggesting that the six lowest energy levels at $\Gamma$ are a direct sum of a two-dimensional irrep and a four-dimensional irrep. The symmetries act on these irreps in the form of the matrices shown in Table~\ref{Tab1}, which have been verified numerically upon proper choice of basis.

\begin{table}[!thb]
\caption{Action of the symmetries $C_3$, $M_y$, $C_2\mathcal{T}$, and $\mathcal{C}$ on the two-dimensional and the four-dimensional irreps of the lowest six bands upon proper choice of basis.}\label{Tab1}
\centering
\begin{tabular}{c|c|c|c}
\hline\hline
Symmetry & \shortstack{Action \\ on 2D irrep} & \shortstack{Action\\ on 4D irrep} & \shortstack{Action\\ on $\bm{k}$}\\
\hline
$C_3$ & $1$ & $e^{\frac{2\pi}{3} i \sigma^z}$ & $\bm{k}\rightarrow C_3(\bm{k})$\\
$M_y$ & $\mu^z$ & $\tau^z\sigma^x$ & $(k_x,k_y)\rightarrow(k_x,-k_y)$\\
$C_2\mathcal{T}$ & $\mathcal{K}$ & $\sigma^x \mathcal{K}$ & $\bm{k}\rightarrow 
\bm{k}$\\
$\mathcal{C}$ & $\mu^x$ & $\tau^x \sigma^x$ & $(k_x,k_y)\rightarrow (-k_x,k_y)$\\
\hline\hline
\end{tabular}
\end{table}

One can then use these symmetry actions to write down the most general forms of effective models for the two irreps, respectively, which are expected to be valid in the vicinity of the $\Gamma$ point. Up to second order, they are given by
\begin{align}\label{eq:bare_effective_hamiltonians}
H_4& = \Delta \, \tau^z + v \, \bm{k}\cdot \bm{\sigma} + \left(d_1k^2+d_2(k^2_+\sigma^+ + k^2_- \sigma^-)\right) \tau^z \nonumber\\
&+ i d_3 (k^2_+\sigma^+- k^2_- \sigma^-) \tau^x + \mathcal{O}(k^3),\nonumber\\
H_2 &= \left( \delta + b_1 k^2 \right) \mu^z ,
\end{align}
where $H_4$ and $H_2$ correspond to the 4D and 2D irreps, respectively, and $k_{\pm}=k_x \pm i k_y$.

In order to capture the low energy physics in the vicinity of $\Gamma$, and in particular, the transitions at $\alpha_1$ and $\alpha_2$, we need a description of the lowest six bands altogether by combining $H_4$ and $H_2$ to form a six-band Hamiltonian. In doing so, cross coupling terms $H_{42}$ are possible and should be included according to symmetry analysis. Since at $\Gamma$ point the two irreps decouple, $H_{42}$ must vanish for $\bm{k}=0$, and to linear order, one has
\begin{equation}
H_{42} = \left(\begin{array}{cc} g_1  k_+ & i g_2 k_+\\ g_1 k_- & - i g_2 k_-\\
-i g_2 k_+ & g_1 k_+ \\ i g_2 k_- & g_1 k_-\end{array}\right).
\label{eq:H_42}
\end{equation}
Combining all these terms listed above, we can write down the six-band Hamiltonian,
\begin{align}\label{eq:six_band_hamiltonian_matrix}
H_6=\begin{pmatrix} H_4 & H_{42} \\
H_{24} & H_2\end{pmatrix},
\end{align}
where $H_{24}=H_{42}^\dag$. One can then fit this effective model to the CM order by order, which will be the subject of the following subsections. 

Let us examine the six band model above qualitatively around $\alpha=\alpha_1$ and $\alpha =\alpha_2$. In $H_4$, when $\Delta=0$, the gap vanishes and there are two DPs at ${\bm k=0}$ with the same velocity; this corresponds to the transition at $\alpha_2$ (see Fig.~\ref{fig:alpha2_linear_a} and \ref{fig:alpha2_linear_b}). The higher order corrections (coming from $H_4$ itself and $H_{42}$ coupling) become more important as we move away from the $\Gamma$ point, and therefore result in deviations between these two cones. 

$\delta = 0$, on the other hand, corresponds to a quadratic touching of the two bands, and thus we expect $\delta$ close to zero to explain the transition at $\alpha = \alpha_1$ (see Fig.~\ref{fig:alpha1_linear_b}).
Indeed, one can derive an effective two-band Hamiltonian from the above six band model via a Schrieffer-Wolf transformation which will result in the following form for the effective Hamiltonian (see Appendix \ref{sec:SW_two_band} for details and derivaion):
\begin{equation}\label{eq:eff_two_band}
\begin{aligned}
	H_2^{\text{eff}} &= c_0 \; \mathrm{Re}(k^3_+) \\
	&+ \left( \delta + c_{3,2} \; k^2 + c_{3,6} \; \mathrm{Re}(k^6_+) \right) \mu^z \\
	&+ c_{1,6} \; \mathrm{Im}(k^6_+)\;\mu^x,
\end{aligned}
\end{equation}
where terms up to sixth order are kept and only the lowest order for each angular-pseudospin dependence is shown here. Note that the $c$ coefficients can be written in terms of the bare coefficients of the original model (Appendix \ref{sec:SW_two_band}). 
This form can also be derived from symmetry considerations. 

The above effective two-band Hamiltonian is capable of explaining how twelve DPs emerge from the $\Gamma$ point at $\alpha = \alpha_1$. To see this note that for $\delta = 0$, there is a quadratic band touching at $\bm{k}=0$, and for small non-zero $\delta$ there is the possibility that DPs are present close to $\Gamma$ if $\delta + c_{3,2}\; k^2 = c_{1,6} \; \mathrm{Im}(k^6_+) = 0$, neglecting the $c_{3,6} \; \mathrm{Re}(k^6_+) $ term. This can only happen if $\delta$ and $c_{3,2}$ have opposite signs, which indeed happens for $\alpha > \alpha_1$ (see next subsections and Appendix \ref{sec:SW_two_band}). Furthermore, one needs to consider the twelve directions given by the angles $\theta_k = \arctan(k_x/k_y) = n (2\pi/12)$, $n=0,1,...,11$, in order for the $\mathrm{Im}(k^6_+)$ term to vanish. Note that this results in twelve DPs along these directions with alternating positive and negative vorticities. The presence of the term $c_0 \; \mathrm{Re}(k^3_+)$ finally indicates that the DPs along $\pm k_x$ and their $C_3$ counterparts are not pinned to zero energy.

\subsection{Six-band model at linear order}
In this subsection, we study the six-band model at linear order in quasi-momentum and compare the results with the CM around $\Gamma$ point in order fix the coefficients in the effective model. At this order, we have $H_4=\Delta \; \tau^z + v \; \boldsymbol{k} \cdot {\bm \sigma}$, $H_2=\delta \; \mu^z$ and $H_{42}$ as in Eq.\eqref{eq:H_42}. $\Delta$ and $\delta$ determine the gaps at ${\bm k=0}$. If we tune $\delta$ to zero with $\Delta>0$, we have two-fold degeneracy at zero energy at ${\bm k=0}$. Around this point, the energy of the middle two bands are equal to $\pm 2[(g_2^2-g_1^2)/\Delta] k^2$ due to the $H_{42}$ coupling terms. In contrast, if we take $\Delta=0$ and $\delta>0$, we have two Dirac cones around ${\bm k=0}$ with the same velocity $v$. As we move away from ${\bm k=0}$ point, the hybridization with $H_2$ induces the splitting between lower and upper bands. We will show below that these two scenarios correspond to $\alpha=\alpha_1$ and $\alpha=\alpha_2$, respectively.

The parameters in the effective model can be evaluated by comparing to the numerical results  {\it very} close to $\Gamma$ point. As the first step, we determine $\delta$ and $\Delta$, which show the value of the energy exactly at the $\Gamma$ point. As seen in Fig.~\ref{fig:Gamma_alpha}, both $\delta$ and $\Delta$ behave linearly around $\alpha_1$ and $\alpha_2$. Since $\delta$ vanishes at $\alpha_1$ and $\Delta$ vanishes at $\alpha_2$, we have $\Delta=S \; (\alpha_2-\alpha)$ and $\delta=s \; (\alpha-\alpha_1)$,
where $s = 1.13$ and $S = 1.29$. 

For the rest of the parameters, we expect them to be smooth functions of $\alpha$ so that they can be expressed as Taylor series around $\alpha_1$ (or $\alpha_2$). In the regime of small enough $k$'s (with only linear $k$ terms in the expansion of the effective mode), one is able to derive expressions for the energy of the six bands and compare them with the numerical results.  For the range of $\alpha$ we are interested in, we find that the band structure can be reproduced by the effective model if we simply treat all coefficients other than $\delta$ and $\Delta$ as constants. Numerically, we find the coefficients of linear order terms to  be: $v = 0.263$, $g_2 = 0.339$ and $g_1 = g_2+0.00146$. Note that we have not attempted to derive the errors corresponding to the above values. Details on how these parameters are derived can be found in Appendix \ref{sec:parameter_six_band}.

\begin{figure}[hbt]
\centering
\subfigure[]{\label{fig:alpha1_linear_a} \includegraphics[width=.23\textwidth]{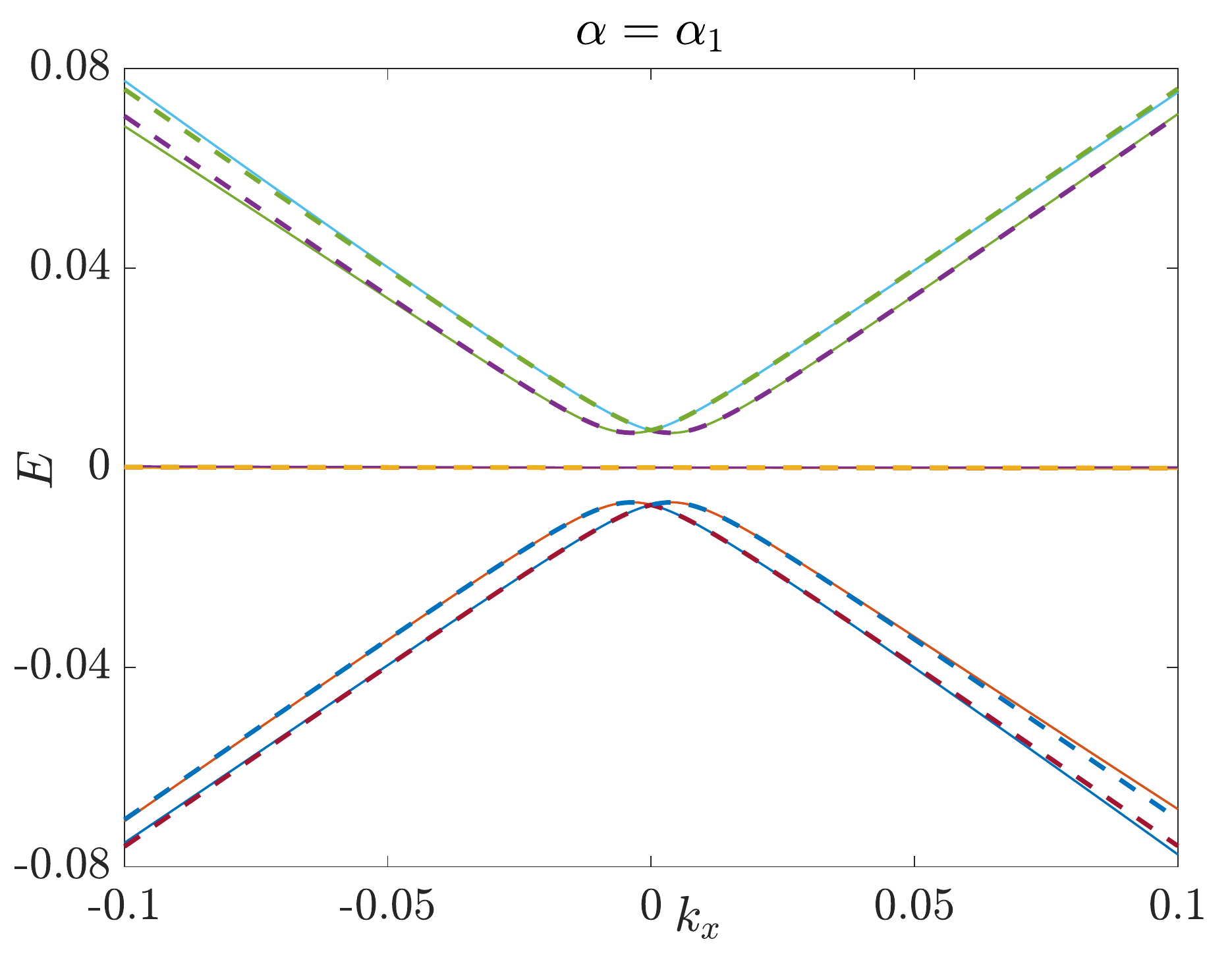}}
\subfigure[]{\label{fig:alpha1_linear_b} \includegraphics[width=.23\textwidth]{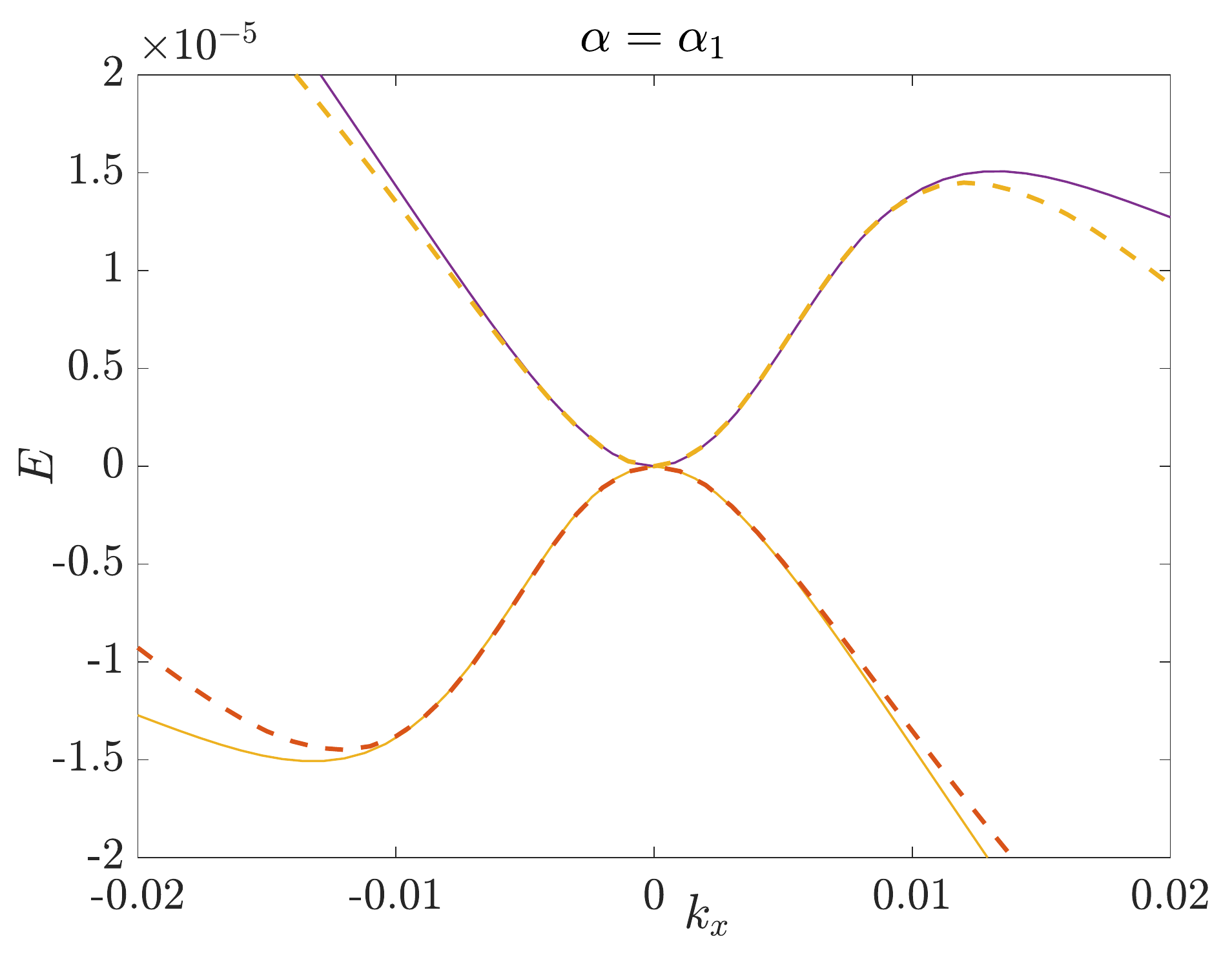}}
 \subfigure[]{\label{fig:alpha2_linear_a} \includegraphics[width=.23\textwidth]{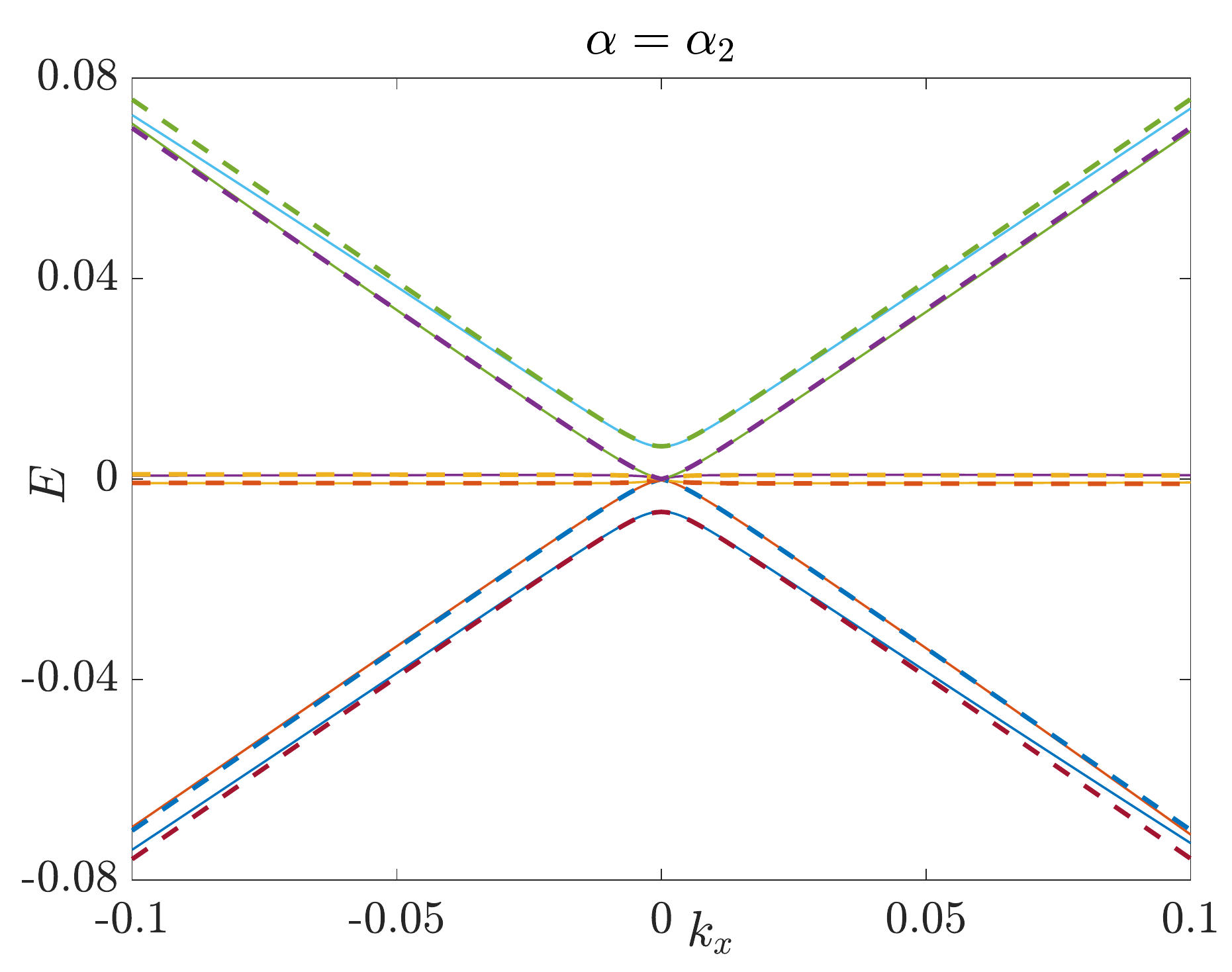}}
 \subfigure[]{\label{fig:alpha2_linear_b} \includegraphics[width=.23\textwidth]{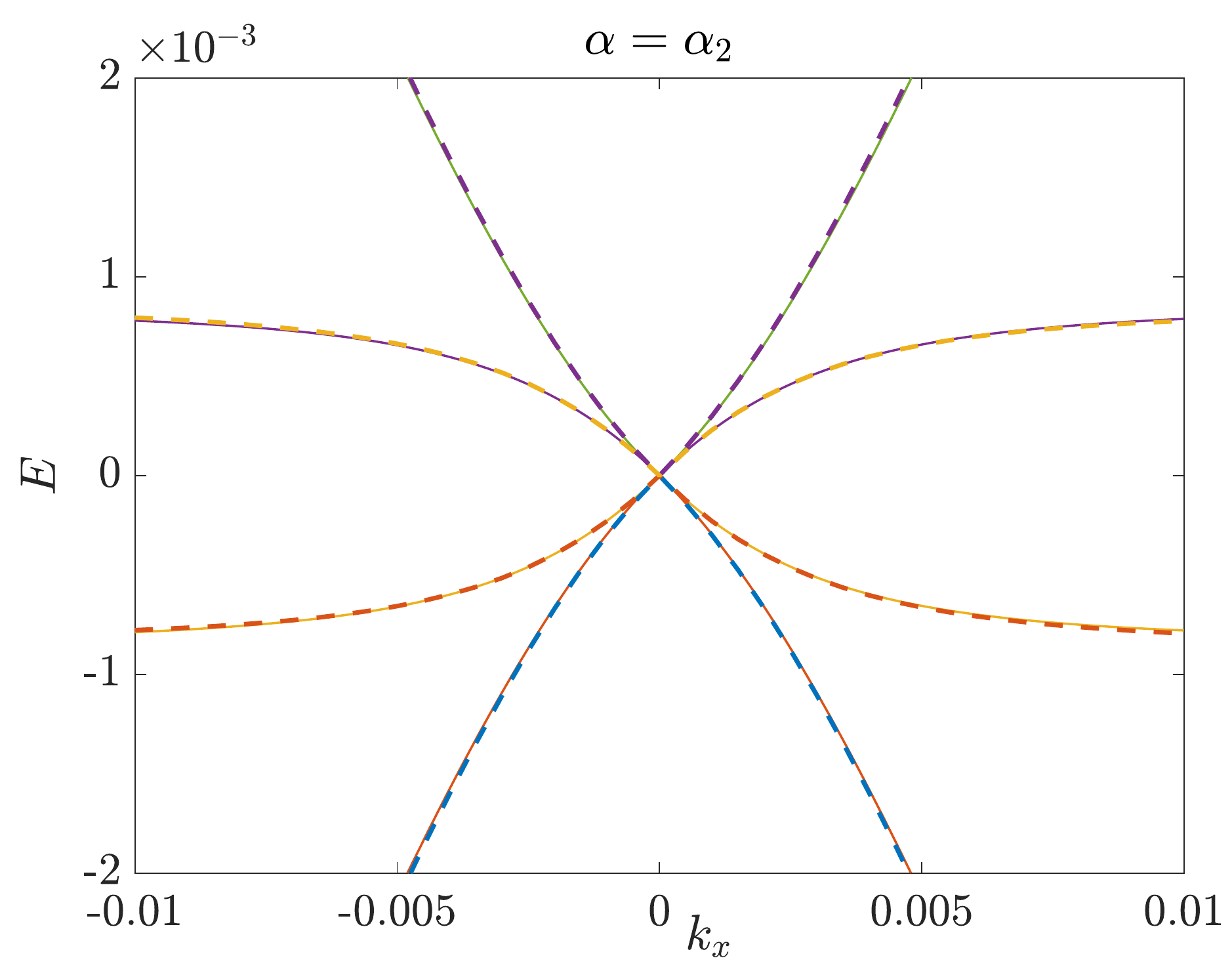}}
\caption{The comparison of energy spectrums between six-band  model at linear order and numerical solution of the CM around $\Gamma$ point. The solid lines are the numerical data and the dashed lines are the results of the six-band model. In all of the plots, $k_y$ is fixed to be zero. (a) is the energy spectrum at $\alpha_1 = 0.56945$ along $k_x$ direction. The middle two bands are very flat and have energy close to zero. (b) is the zoomed-in structure of the two middle bands in a smaller range of $k_x$. These two bands touch at $k_x=0$ and have quadratic dispersion around it. (c) is the energy spectrum at $\alpha_2 = 0.57524$ along $k_x$ direction. The middle four bands form two Dirac cones around the $\Gamma$ point with the same velocity as is seen in (d). However they show deviations for larger values of $|\bm{k}|$; interestingly for $|\bm{k}|>0.1$, the two middle bands are very flat with energy  close to zero.} 
\label{fig:six_band_linear_k_x}
\end{figure}

\begin{figure}[hbt]
\centering
 \subfigure[]{\label{fig:alpha1_Delta_E_a} \includegraphics[width=.23\textwidth]{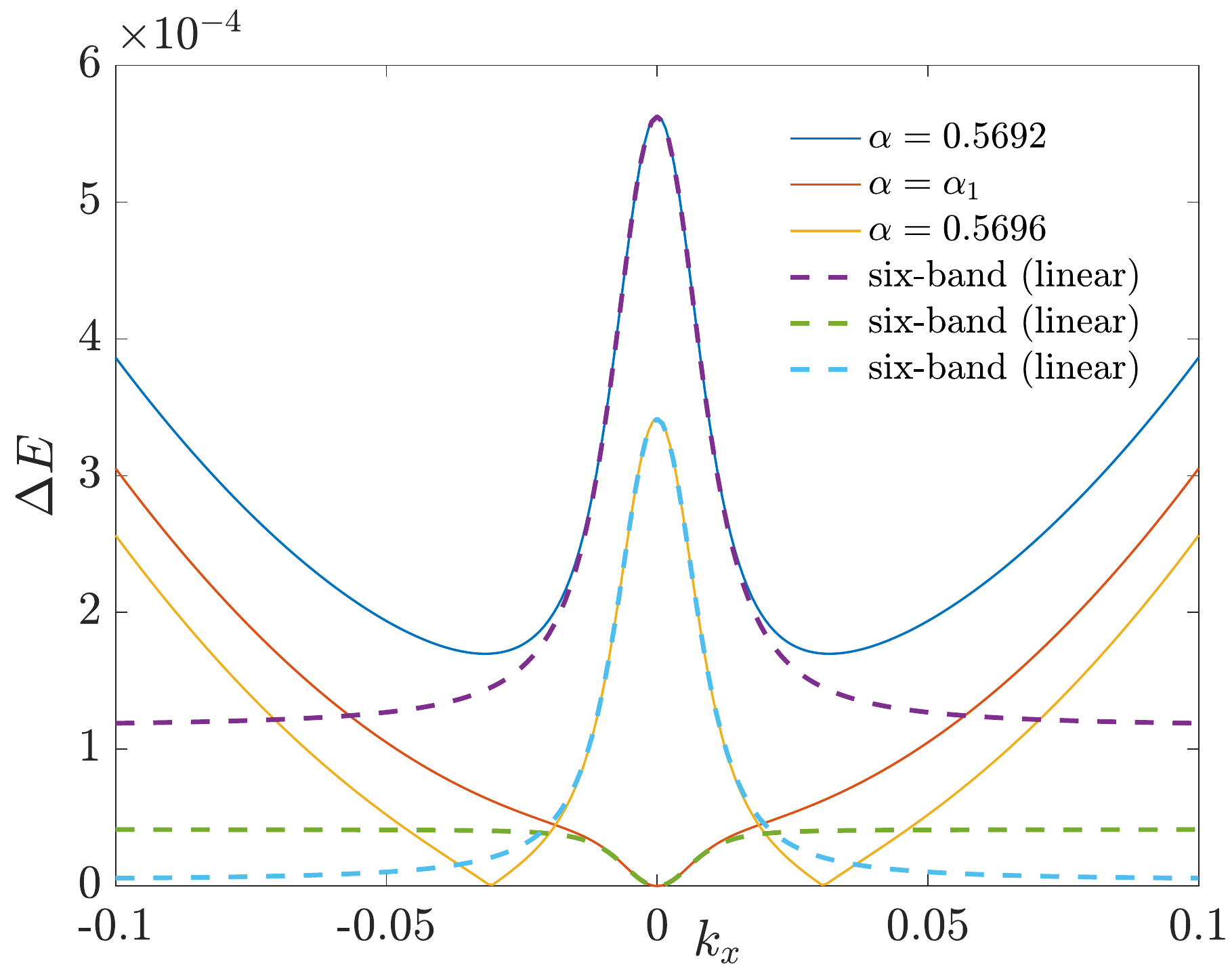}}
 \subfigure[]{\label{fig:alpha1_Delta_E_b} \includegraphics[width=.23\textwidth]{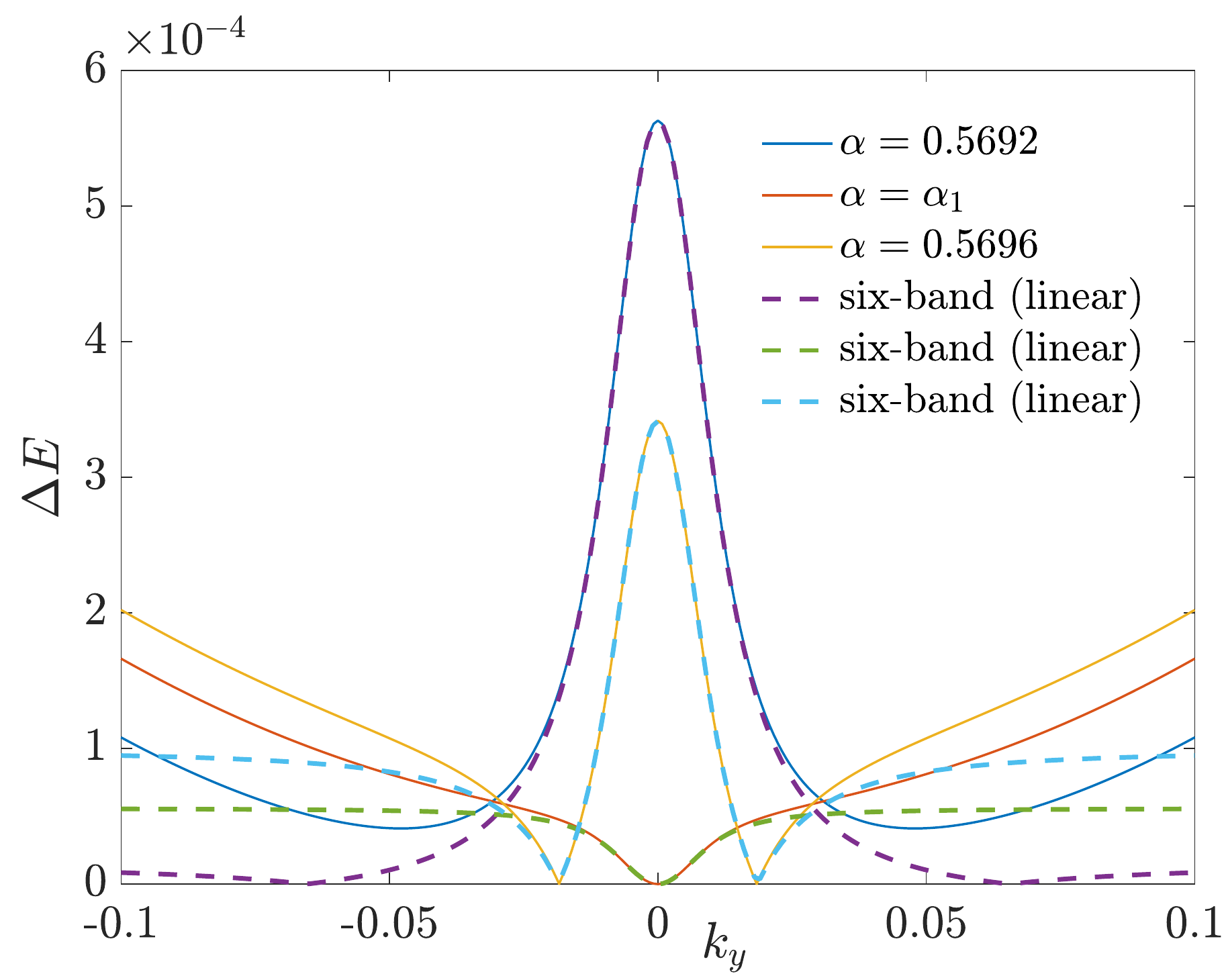}}
 \subfigure[]{\label{fig:alpha2_Delta_E_a} \includegraphics[width=.23\textwidth]{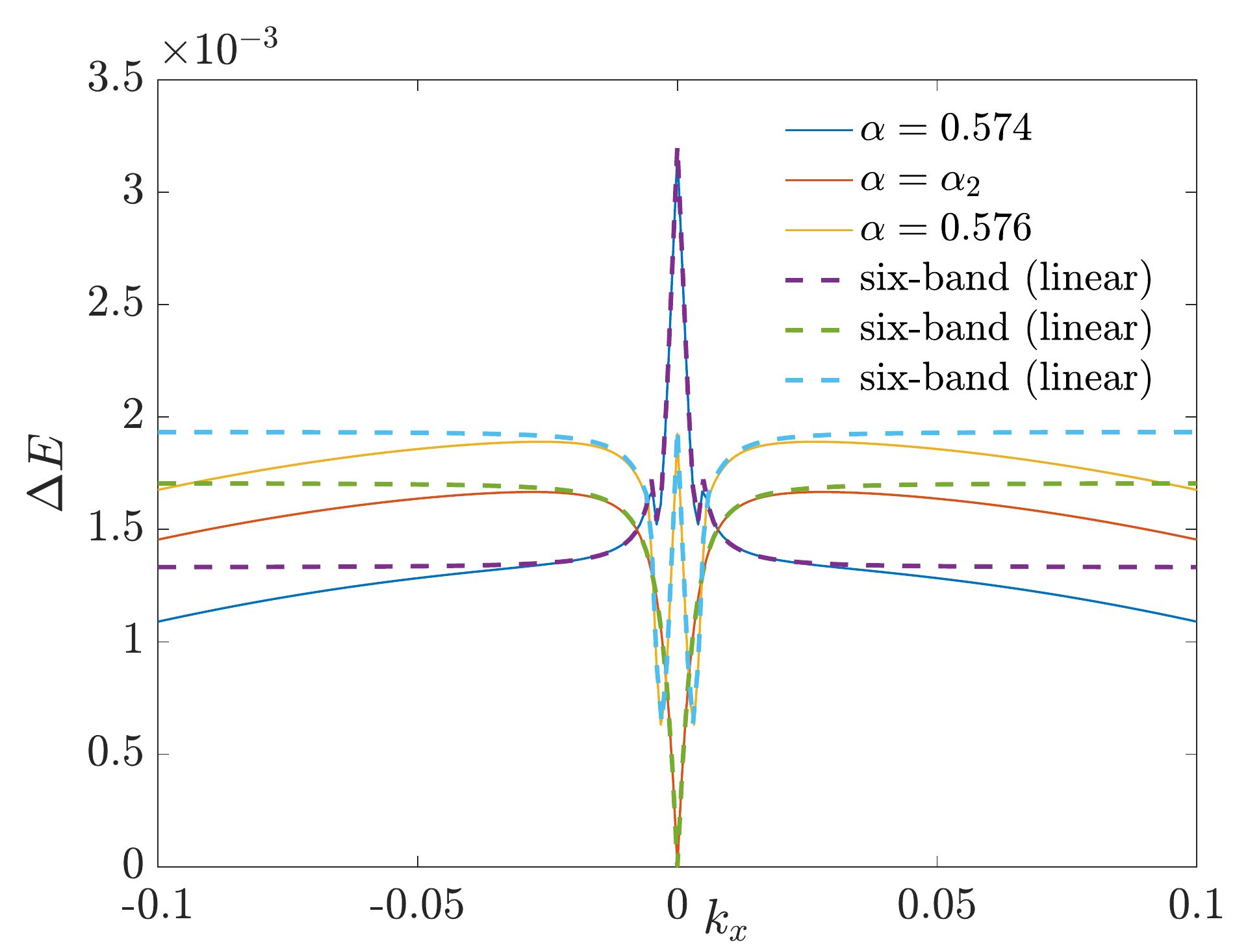}}
 \subfigure[]{\label{fig:alpha2_Delta_E_b} \includegraphics[width=.23\textwidth]{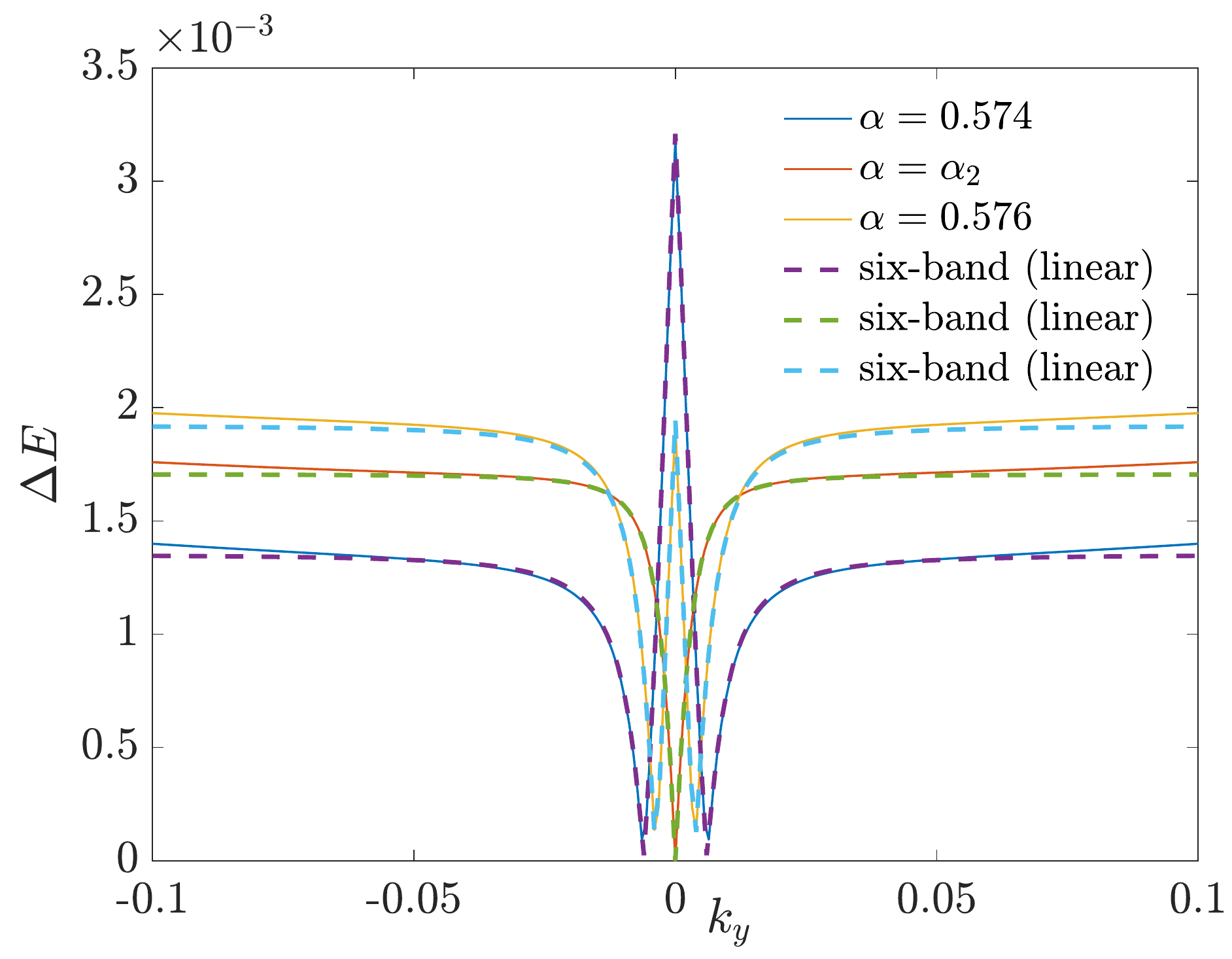}}
\caption{ The comparison of $\Delta E$ of the middle two bands between six-band model at linear order and numerical data from CM along $k_x$ and $k_y$ directions. The solid lines are the numerical data and the dashed lines are the results from the six-band model. (a) is the result around $\alpha_1 = 0.56945$ along $k_x$ direction with fixed $k_y=0$. (b) is the result along $k_y$ direction with fixed $k_x=0$. (c) is the result around $\alpha_2 = 0.57524$ along $k_x$ direction with fixed $k_y=0$. (d) is the result along $k_y$ direction with fixed $k_x=0$.} 
\label{fig:six_band_linear}
\end{figure}

In Fig.~\ref{fig:six_band_linear_k_x}, we compute the six bands closest to zero energy in the CM and compare them with the six-band model. At $\alpha=\alpha_1$, we see that the middle two bands are well separated from the rest of the four bands, and that there is a very good agreement between the computed middle two bands and the effective model, when $k_x\in[-0.015, 0.015]$. The quadratic dispersion of these two bands around $\Gamma$ point is caused by the coupling with the other four bands. On the other hand, at $\alpha=\alpha_2$, those other four bands have lower energy and form two Dirac cones in the vicinity of $\Gamma$ point with the same velocity $v$. Here, again the agreement is very good within the interval mentioned above.

In Fig.~\ref{fig:six_band_linear}, we present $\Delta E=E_+-E_-$ of the middle two bands for various values of $\alpha$ and compare them with the six-band model. The results agree for $|\bm{k}| \lessapprox 0.02$  in different plots, however, we notice qualitative deviations at larger momentum. At $\alpha=\alpha_1$, the CM results show a $k^2$ behavior for larger $|\bm{k}|$, though with a different coefficient than the similar $k^2$ behavior that occurs for very small $|\bm{k}|$. In contrast, in the effective model, $\Delta E$ saturates to some constant in both $k_x,k_y$ directions and therefore cannot reproduce the correct physics at large momenta.

With all this said, we should note that the six-band model up to linear order in $k$
provides a faithful approximation of the band structure
for $|\bm{k}| \lessapprox 0.02$ as long as $\alpha$ is close enough to $\alpha_1$ and $\alpha_2$. The discrepancy beyond this regime can be resolved by considering the quadratic corrections in the effective Hamiltonian, which is the subject of the next subsection.

\subsection{Quadratic correction}

In this subsection, we study the effects of quadratic corrections in the effective model. The effective six-band model up to second order in $k$ now contains $H_2$ and $H_4$ as given in Eq.~\eqref{eq:bare_effective_hamiltonians}, along with $H_{42}$ as in Eq.~\eqref{eq:H_42}. 

Our goal, similar to what was done in the previous subsection, is to find the coefficients in the second order effective Hamiltonian by fitting to well-chosen combinations of energies in the CM. In particular, we use the information from the second quadratic behavior discussed above in determining the coefficients. We show in Appendix \ref{sec:parameter_six_band} how the following set of coefficients are derived:
\begin{eqnarray}
		&&v = 0.263, \qquad g_2 = 0.339, \qquad g_1 - g_2 = 0.00130, \nonumber\\
		&&b_1 = -0.0289, \qquad d_1 = 0, \qquad  d_2 = -0.0106 \qquad d_3 = 0.\nonumber\\
\end{eqnarray}
As one can see, the values found above for the linear coefficients are slightly different from those found when only linear terms were taken into account.

\begin{figure*}[hbt]
\centering
 \subfigure[]{\label{fig:alpha_E_05688} \includegraphics[width=.25\textwidth]{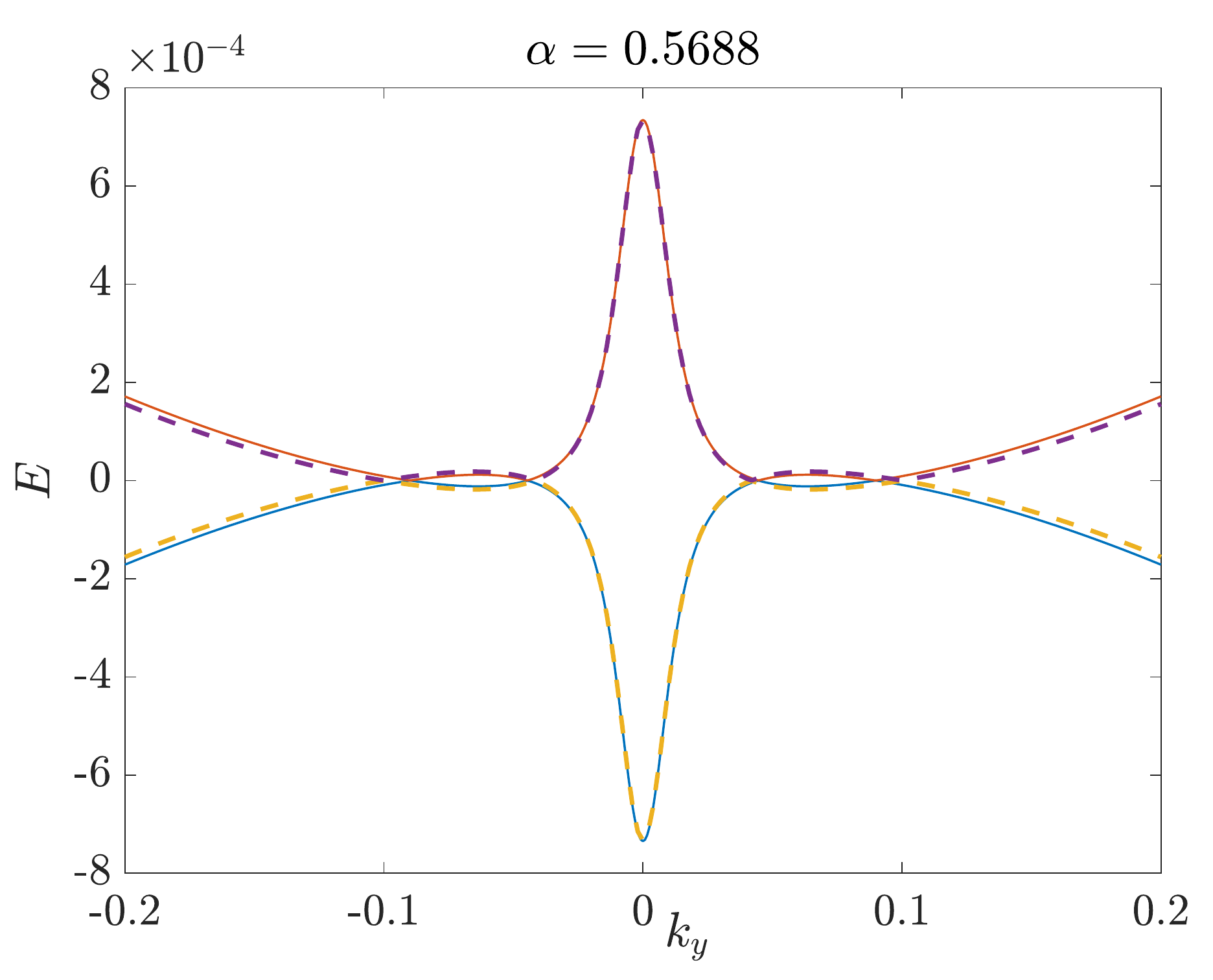}}\hspace{0.5cm}
 \subfigure[]{\label{fig:alpha_E_056895} \includegraphics[width=.25\textwidth]{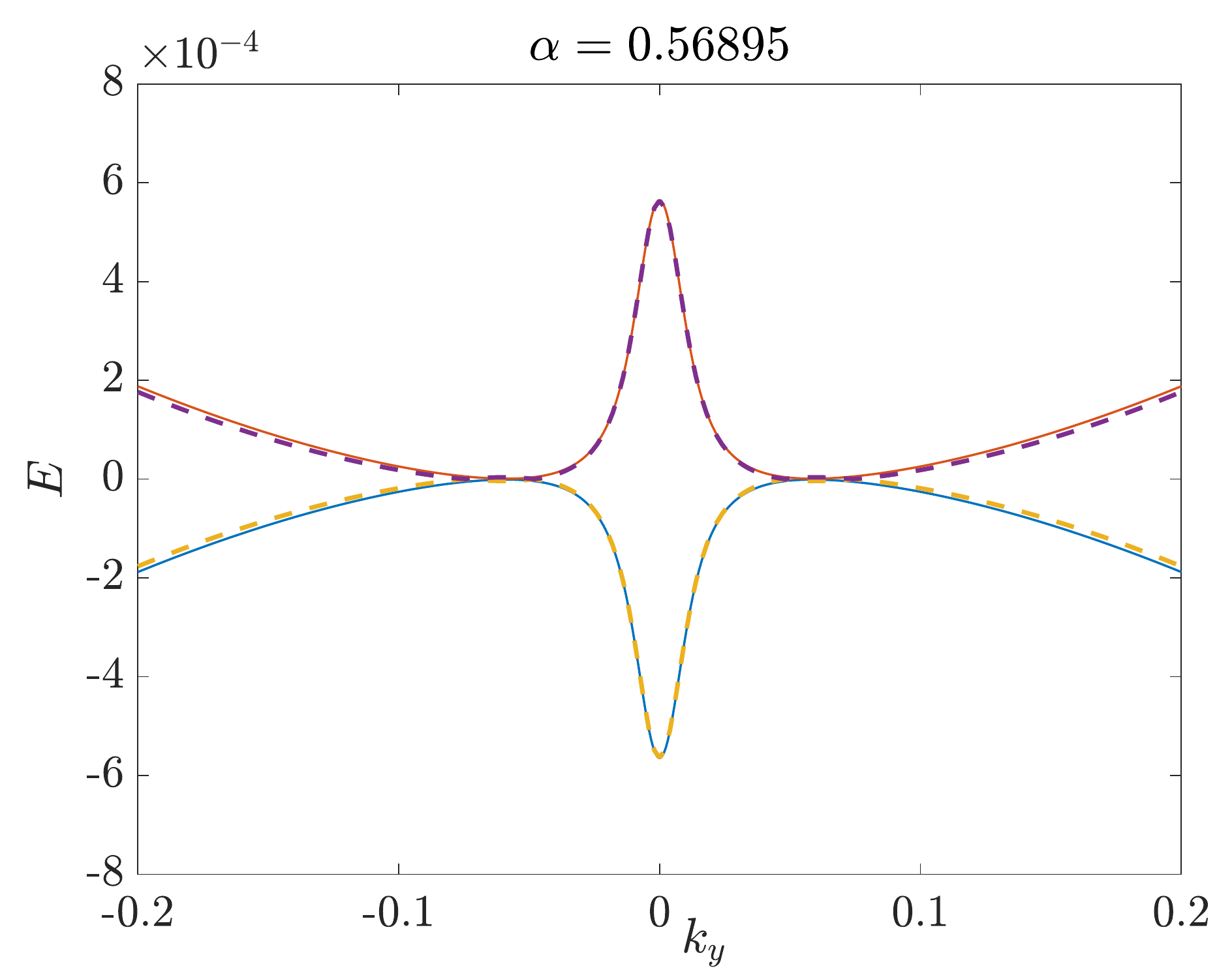}}\hspace{0.5cm}
 \subfigure[]{\label{fig:alpha_E_05692} \includegraphics[width=.25\textwidth]{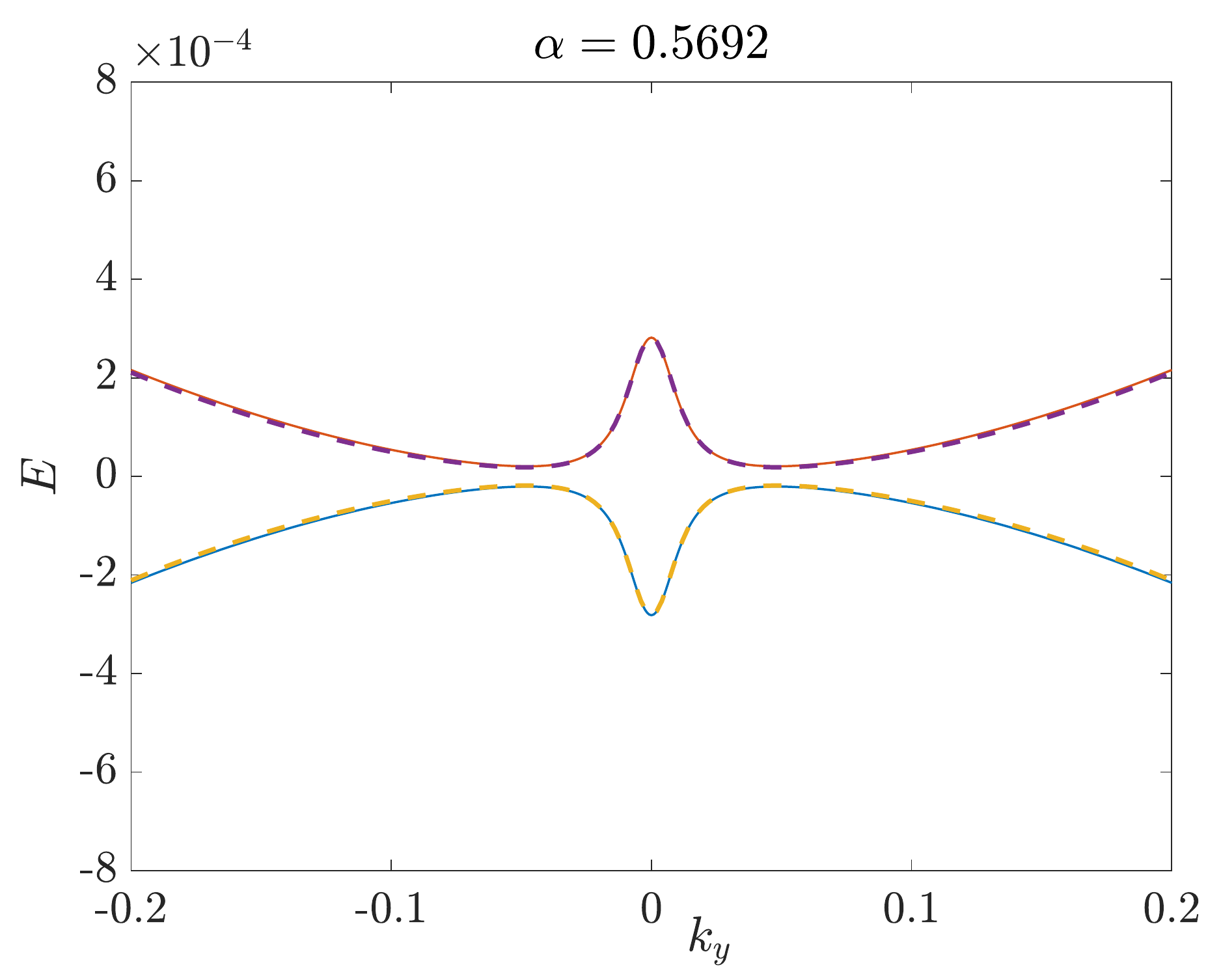}}
 \subfigure[]{\label{fig:alpha_E_056945} \includegraphics[width=.25\textwidth]{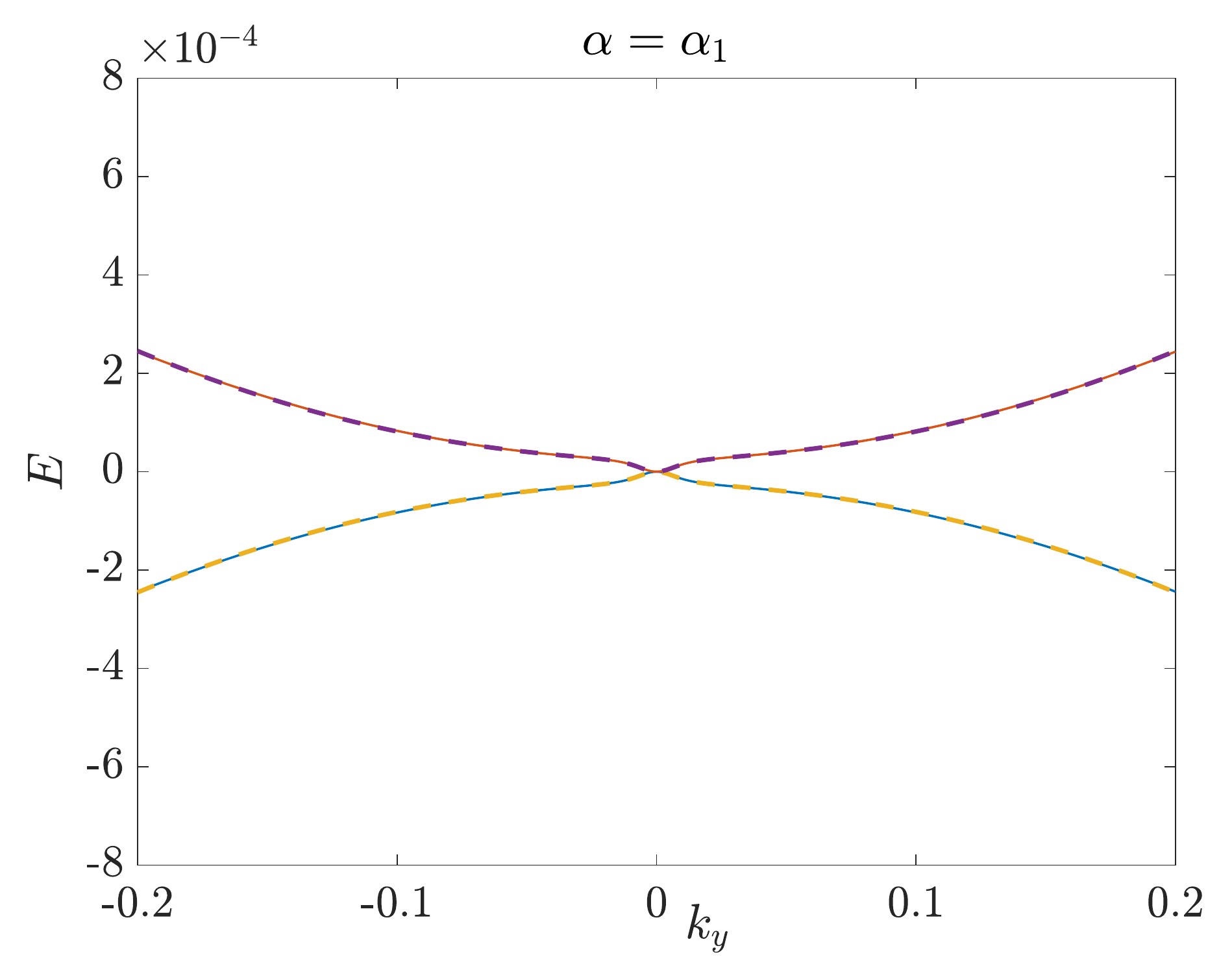}}\hspace{0.5cm}
  \subfigure[]{\label{fig:alpha_E_05697} \includegraphics[width=.25\textwidth]{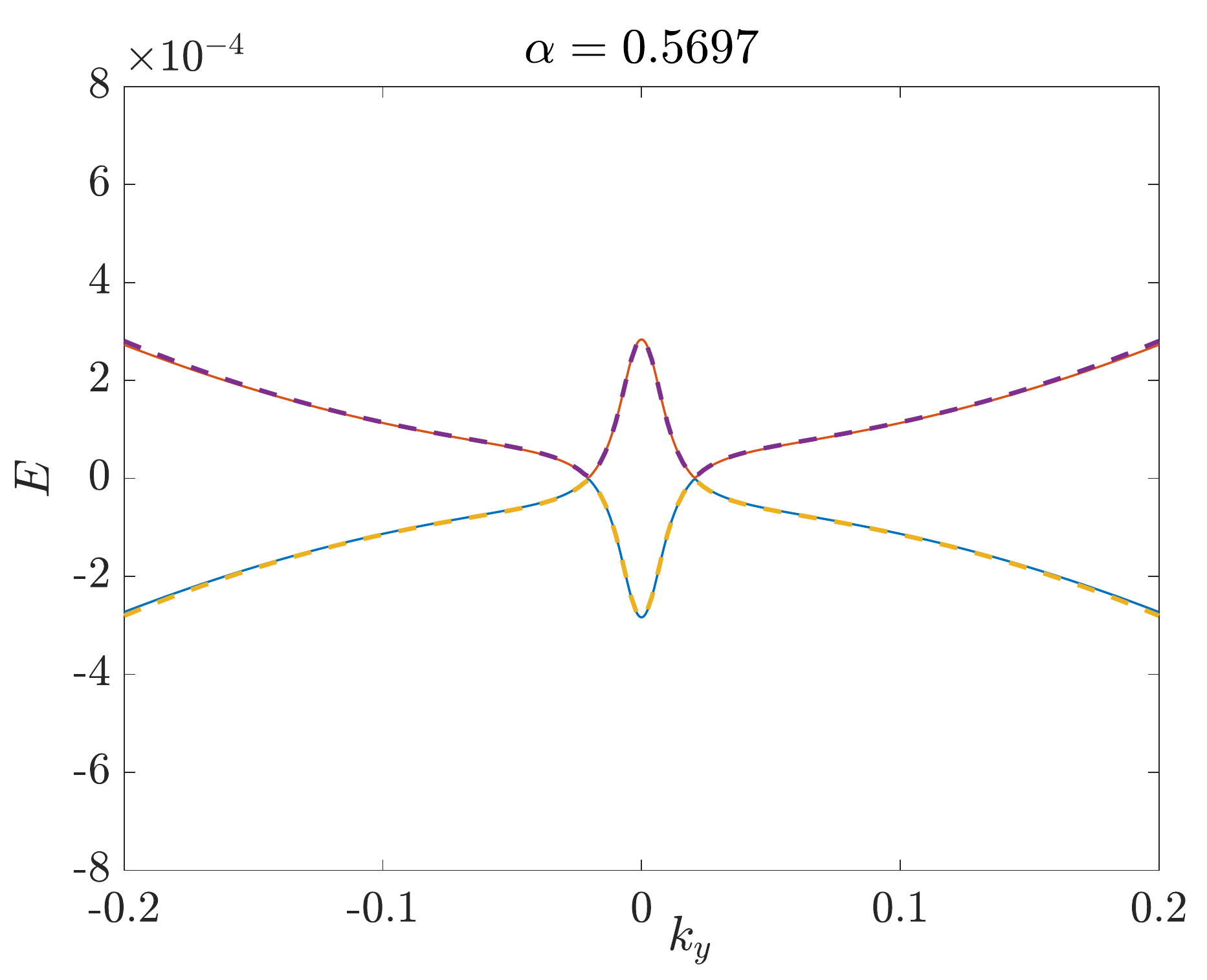}}\hspace{0.5cm}
   \subfigure[]{\label{fig:alpha_E_057} \includegraphics[width=.25\textwidth]{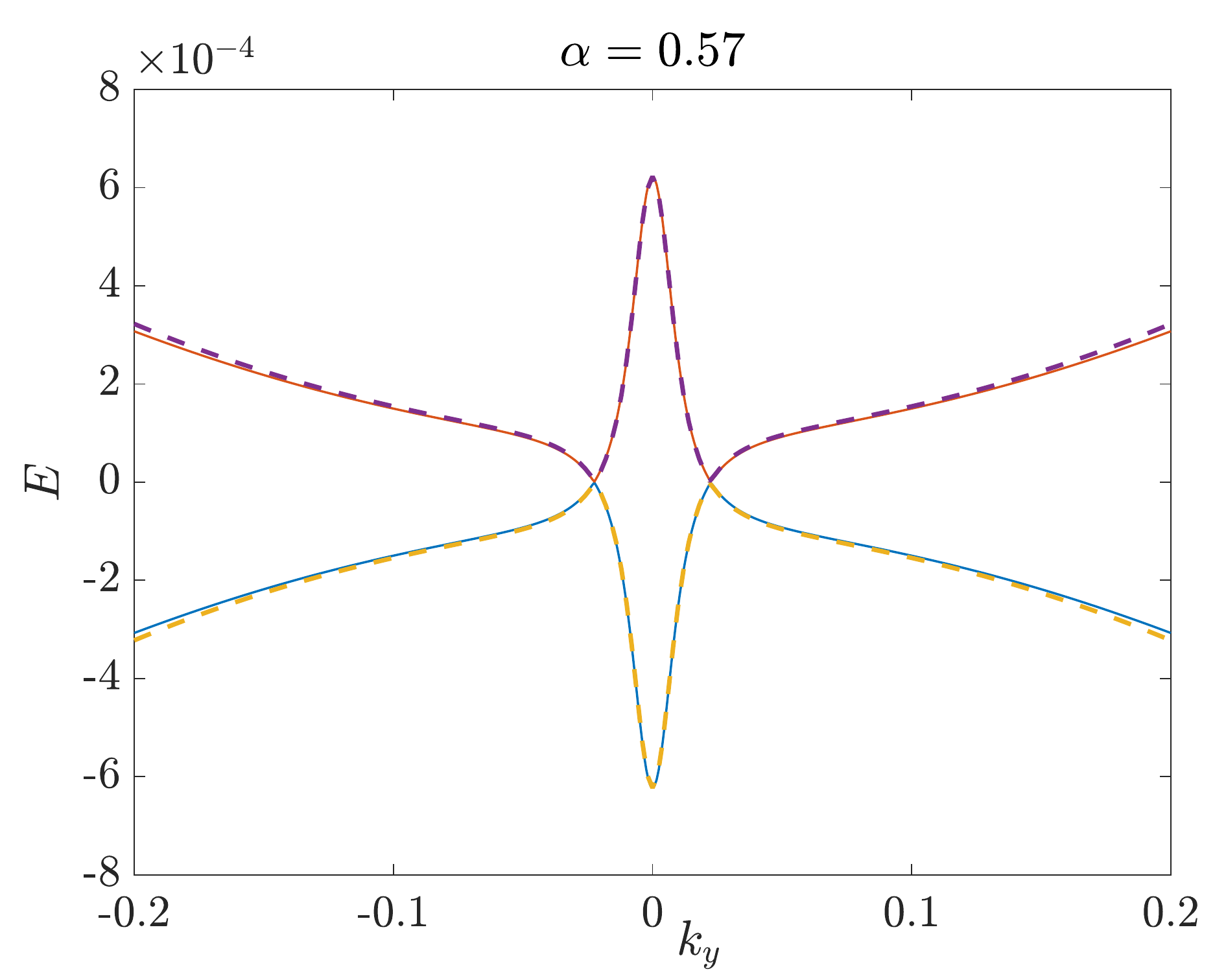}}
   \subfigure[]{\label{fig:alpha_E_05688_kx} \includegraphics[width=.25\textwidth]{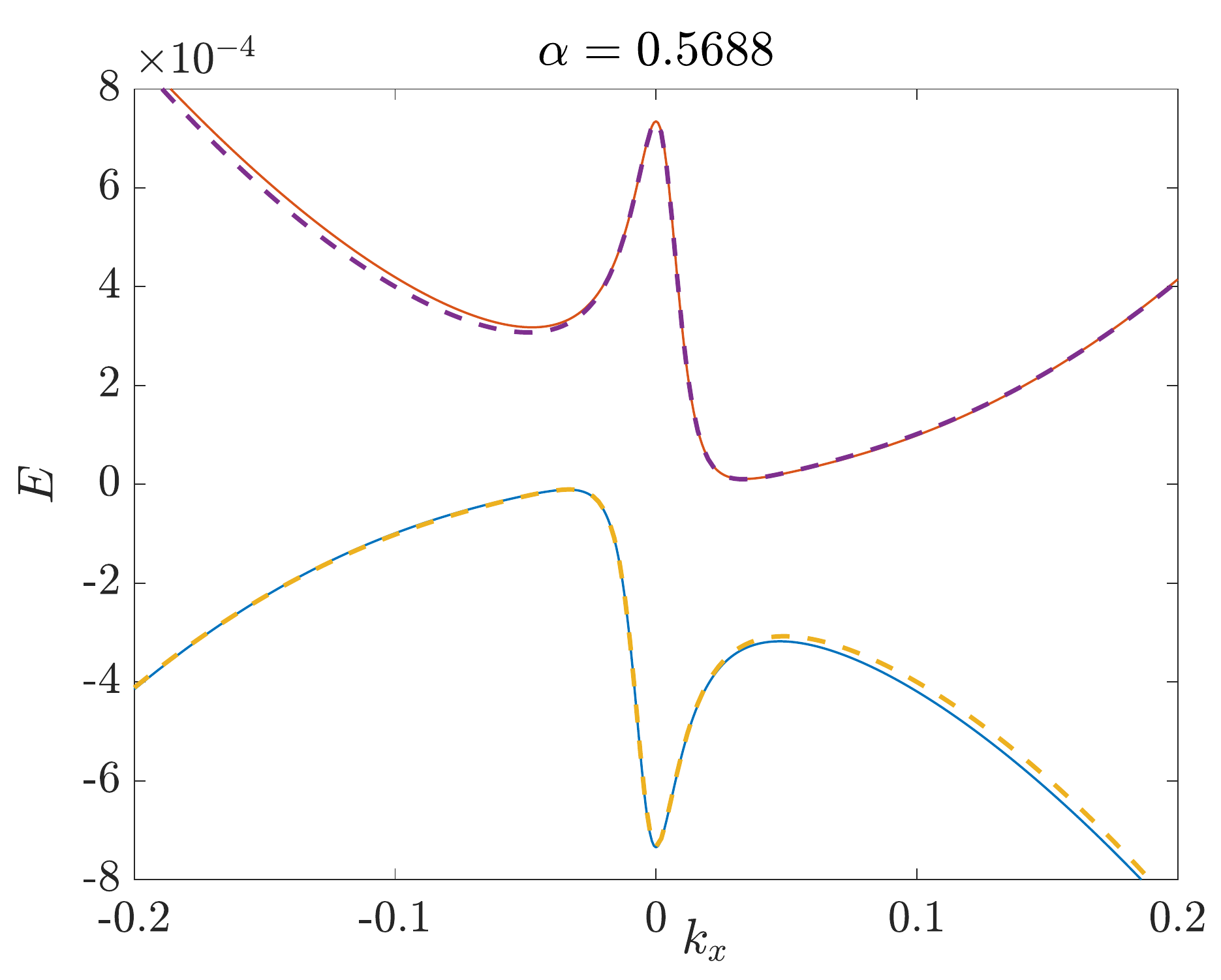}}\hspace{0.5cm}
  \subfigure[]{\label{fig:alpha_E_056945_kx} \includegraphics[width=.25\textwidth]{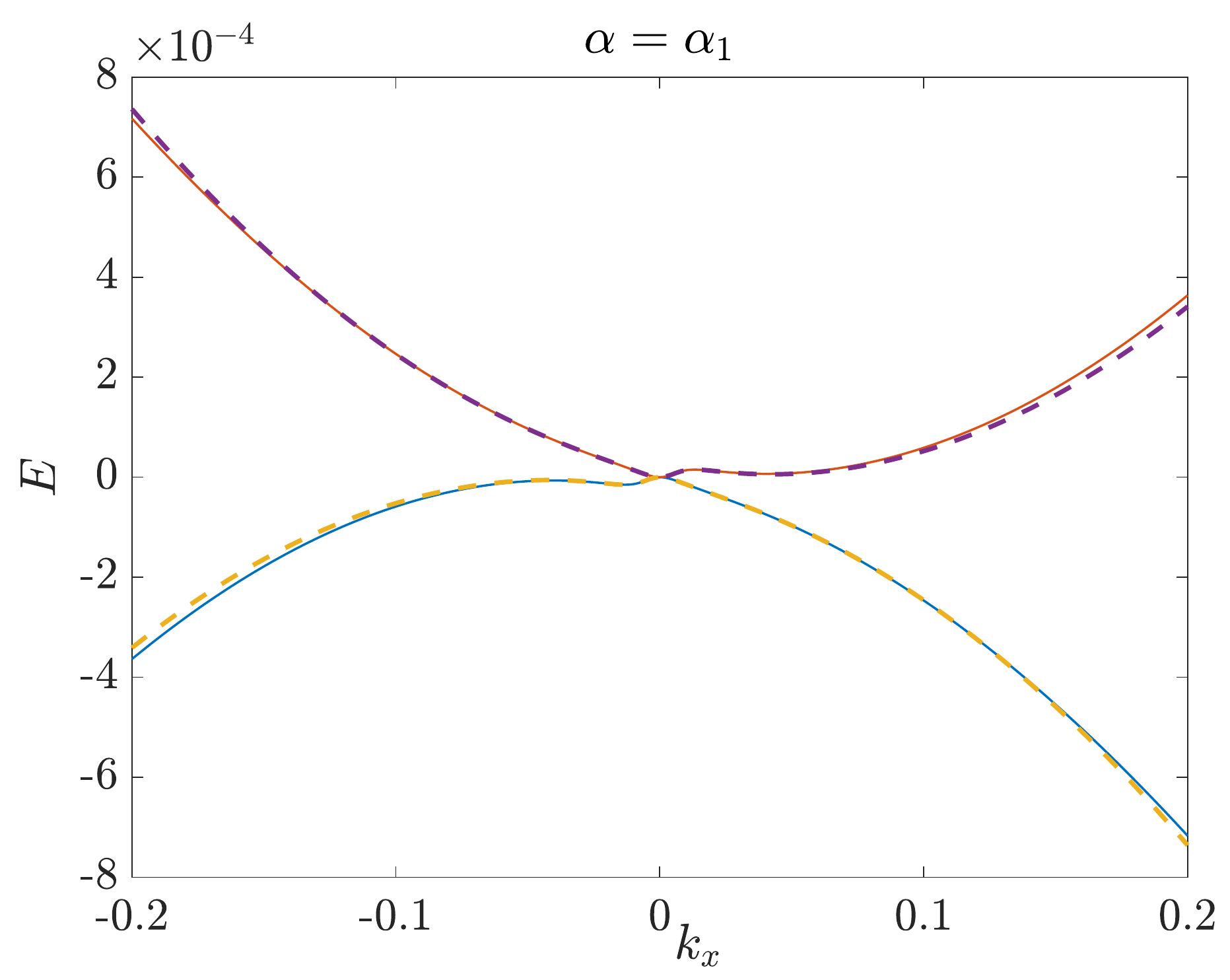}}\hspace{0.5cm}
   \subfigure[]{\label{fig:alpha_E_057_kx} \includegraphics[width=.25\textwidth]{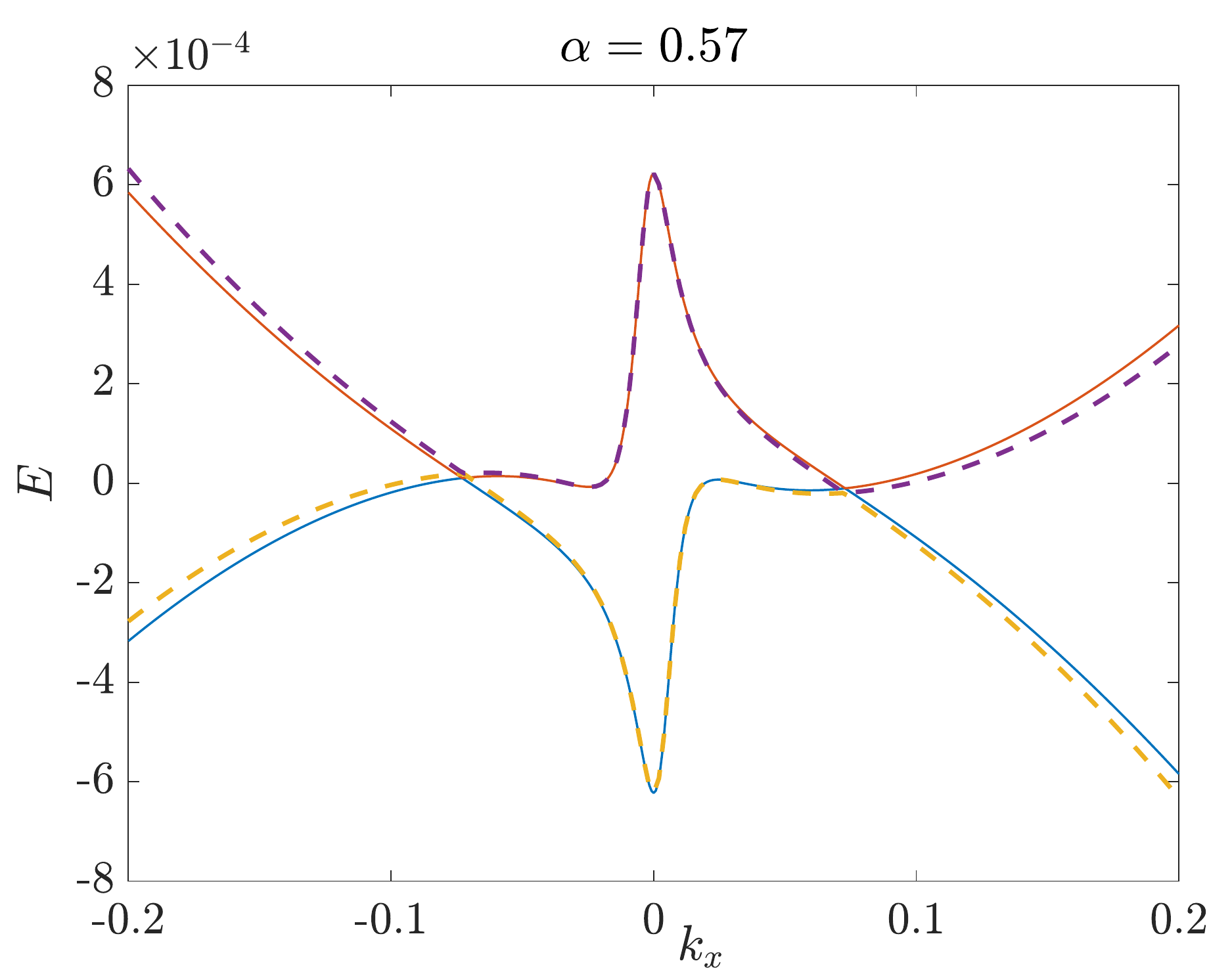}}
\caption{ The energy of the middle two bands closest to zero energy along $k_y$ or $k_x$ directions at various $\alpha$ values (near $\alpha_1 = 0.56945$). The solid lines are the numerical data and the dashed lines are the results from the six-band model with quadratic correction. In (a)-(f), $k_x$ is fixed to be zero, while in (g)-(i), $k_y$ is fixed to be zero.} 
\label{fig:six_band_quadratic}
\end{figure*}

Figure~\ref{fig:six_band_quadratic} shows a comparison between these results and the CM, where the two middle energy bands are plotted when $\boldsymbol{k}$ lies in the $k_y$ or $k_x$ direction; it is evident that the agreement between the two results has now expanded to a much wider interval around the $\Gamma$ point, for a variety of $\alpha$'s chosen close to $\alpha_1$. Notice that around $\alpha_2$, we observe similar improvement as we include second order terms, however, since the number and location of the 6 DPs are already correctly described by the six-band model at linear order, we focus on $\alpha$ near $\alpha_1$ in this subsection.

It is now worthwhile to discuss the location of the DPs of the two middle bands, and see how they are captured by the effective six-band model with quadratic correction. First,  when $\alpha$ is slightly larger than $\alpha_1$, twelve DPs emerge from the $\Gamma$ point as shown in Fig.~\ref{fig:schematic_R}. This emergence of the DPs is captured by the effective model even at the linear level. However, as we increase $\alpha$ further, the DPs along $\pm k_x$ direction and their $C_3$ counterparts start to move apart from the $\Gamma$ point (Fig. \ref{fig:schematic_R}) and cannot be captured by the six-band model at the linear order. Thus, in order to reproduce these six moving DPs in the effective Hamiltonian, the quadratic corrections are required (see Fig.~\ref{fig:alpha_E_057_kx}). In contrast, the linear-order effective Hamiltonian is sufficient to reproduce the other six almost stationary DPs along $\pm y$ direction and their $C_3$ counterparts over the parameter range $\alpha_1 \leq \alpha \leq \alpha_2$ and even beyond; the reason is that these DPs always stay close to $\Gamma$  (recall Fig.~\ref{fig:schematic_R}). Also, the effective model replicates the accidental non-topological crossings along $\pm k_x$ and the $C_3$ equivalent directions at values of $\alpha$ very close to $\alpha_2$.

The inclusion of the second order term in the effective Hamiltonian is also crucial for describing the band structure in the parameter range $\alpha<\alpha_1$ (upper panel of Fig.~\ref{fig:schematic_R}): as $\alpha$ is decreased below $\alpha_1$, the DPs first disappear  and a small gap forms between these two bands (Fig.~\ref{fig:alpha_E_05692}). At $\alpha=0.56895$, the gap closes and these two bands touch quadratically along the $\pm k_y$ and the other $C_3$ equivalent directions (Fig.~\ref{fig:alpha_E_056895}). If $\alpha$ is decreased further, twelve DPs emerge along the $\pm k_y$ and its $C_3$ equivalent directions (Fig.~\ref{fig:alpha_E_05688}). Note that there is no band touching along the $\pm k_x$ (and its $C_3$ equivalent directions) in this regime (Fig.~\ref{fig:alpha_E_05688_kx}). 
 
Based on the above analysis, we observe that adding the quadratic terms to the effective Hamiltonian lets us reproduce most features of the CM within a larger window of momenta $|\bm{k}| \lessapprox 0.2$ around $\Gamma$ over the range of interest of $\alpha$.

\section{Conclusion and Outlook}\label{sec:conclusion_outlook}

In summary, we have presented a comprehensive study of the CM of twisted bilayer graphene, focusing on how the middle two bands in this model (the celebrated ``flat bands'') evolve upon changing the twist angle around the first magic angle. We have found that close to the first magic angle, there is actually a range of $\alpha$ in which two bands are nearly flat. Multiple topological transitions occur in the  BZ over this range of $\alpha$.

As a first step, we discussed the Lifshitz transitions at $\text{K}$ and $\text{K}^\prime$ points around $\alpha=\alpha_0=0.6051$, which is denoted as the ``first magic angle'' in the literature. We have shown that there is always a two-fold degeneracy at $\text{K}$ and $\text{K}^\prime$ points due to $C_2\mathcal{T}$ and $C_3$ symmetries, and it is locked to zero energy by the particle-hole symmetry. At $\alpha = \alpha_0$, the two middle bands of the CM touch quadratically at $\text{K}$ and $\text{K}^\prime$ points. As we move away from $\alpha_0$, each of these quadratic band touching points breaks into four DPs, with one DP always at $\text{K}$ or $\text{K}^\prime$ point surrounded by three DPs with opposite topological charge. We kept track of the motion of these three DPs and found that they  all originate from the $\Gamma$ point, which is the highest symmetry point in the BZ. 

This finding motivates us to study the band structure in the vicinity of the $\Gamma$ point in detail. We noticed that the band structure near $\Gamma$ point is rather complicated and is very sensitive to small changes in $\alpha$. Roughly speaking, for a given $\alpha$ slightly smaller than $\alpha_0$, twelve DPs emerge from the $\Gamma$ point; six of them with the same positive charge move to the $\text{K}$ and $\text{K}^\prime$ points as $\alpha$ is increased and lead to the Lifshitz transition at $\alpha_0$. The other six with the negative charge stay close to the $\Gamma$ point; at $\alpha=0.74>\alpha_0$, these six DPs are annihilated with the other six positively charged DPs somewhere close to the $\text{K}$ or $\text{K}^\prime$ points; this leaves behind the only two DPs at $\text{K}$ and $\text{K}^\prime$. 

To better understand the rich physics of the topological transitions near the $\Gamma$ point, we proposed an effective six-band model which incorporates all the symmetries satisfied at $\Gamma$. We demonstrated that when its parameters are carefully tuned with reference to the solution of the CM, this  six-band model is capable of capturing the essential physics that happens close to the $\Gamma$ point for the range of $\alpha$ we are interested in. In particular, the six-band model, which includes both linear and second order terms in quasi-momentum $\boldsymbol{ k}$, predicts the correct number and locations of DPs. This effective six-band model is much simpler than the CM and can be a starting point to study strongly correlated phenomena as we add four fermion interactions.

One interesting question to ask is: why are the bands so flat near the magic angles? One possible answer is that there are in fact many DPs required to exist near these angles, and this fact forces the two bands to repeatedly approach each other at many points in the BZ. To better illustrate this point, we further studied the band structure near the second and third magic angles (results are not shown in this manuscript) and we observed that very similar physics, although not identical, happens over there. The Lifshitz transitions at the $\text{K}$ and $\text{K}^\prime$ points are still the same, while the topological transitions around $\Gamma$ point involve more DPs and are more complicated.

One potential future research direction is to study modified versions of the CM in which some of the symmetries are broken. The CM is a highly symmetric model; the much more complicated experimental TBG system, however, is not expected to preserve all the symmetries of the CM. As a better simulation of the real system, it would be interesting to carry out the stability analysis of the DPs upon including symmetry breaking terms in the CM from physical considerations.
For instance, particle-hole symmetry of the CM is broken by various effects: it is artificial and its breaking allows DPs to move away from zero energy. We have remarked that the $\theta$ rotation for the sublattice matrices in the Dirac term breaks this symmetry; it would be interesting to study how this modifies topological transitions discussed in this work. Preliminary results indicate that adding the $\theta$ rotation will result in some modifications indeed: first, the second gapless point at $\Gamma$ namely the one that occurs at $\alpha = \alpha_2$ disappears. However, one can check that the $\Gamma$ point transition at $\alpha = \alpha_1$ and the transitions at $\mathrm{K}$ and $\mathrm{K}'$ points at $\alpha = \alpha_0$ are all preserved. The DP transfers sketched in Fig.~\ref{fig:Dirac_motion} are still qualitatively true, although altered slightly in details. We will present a thorough study of these effect in a future work.

 Another very important process is the lattice relaxation~\cite{nam2017lattice} of the two layers of graphene, which is not taken into account in the CM, while it can be the dominant source of particle-hole symmetry breaking. In short, the in-plane relaxation process leads to a deformed lattice with larger AB/BA stacking regions, since the AB/BA stackings are more energetically favored.  It is worth studying the topological phase transitions in presence of lattice relaxation and examining the strongly correlated physics including this effect. Furthermore, the out-of-plane lattice relaxation can also significantly modify the low energy bands, for instance it can increase the gap between the middle and the higher bands, which can lead to a more experimentally relevant regime. A systematic study of the CM in presence of lattice relaxations can be the subject of a future work.

\emph{Note Added--}Recently, another paper \cite{2018arXiv180710676S} studied the topological transitions and representation of the low energy bands of the CM, which is consistent with our work.

\begin{acknowledgements}

X.C. was supported by a postdoctoral fellowship from the the Gordon and Betty Moore Foundation,
under the EPiQS initiative, Grant GBMF4304, at the Kavli Institute for Theoretical
Physics. 
H.S. acknowledges the support from the National Science Foundation under Grant No.\ DMR-1455296
and the KITP graduate fellowship program (NSF PHY-1748958).
We acknowledge support from the Center for Scientific Computing from the CNSI, MRL: an NSF MRSEC (DMR-1121053).  Research of L.B., C.L. and K. H. was supported by the National Science Foundation grant DMR1506119.

\end{acknowledgements}

\appendix
\section{Geometry of twisted bilayer graphene}
\label{sec:geometry}
\begin{figure}[h]
\centering
\includegraphics[width=0.4\textwidth]{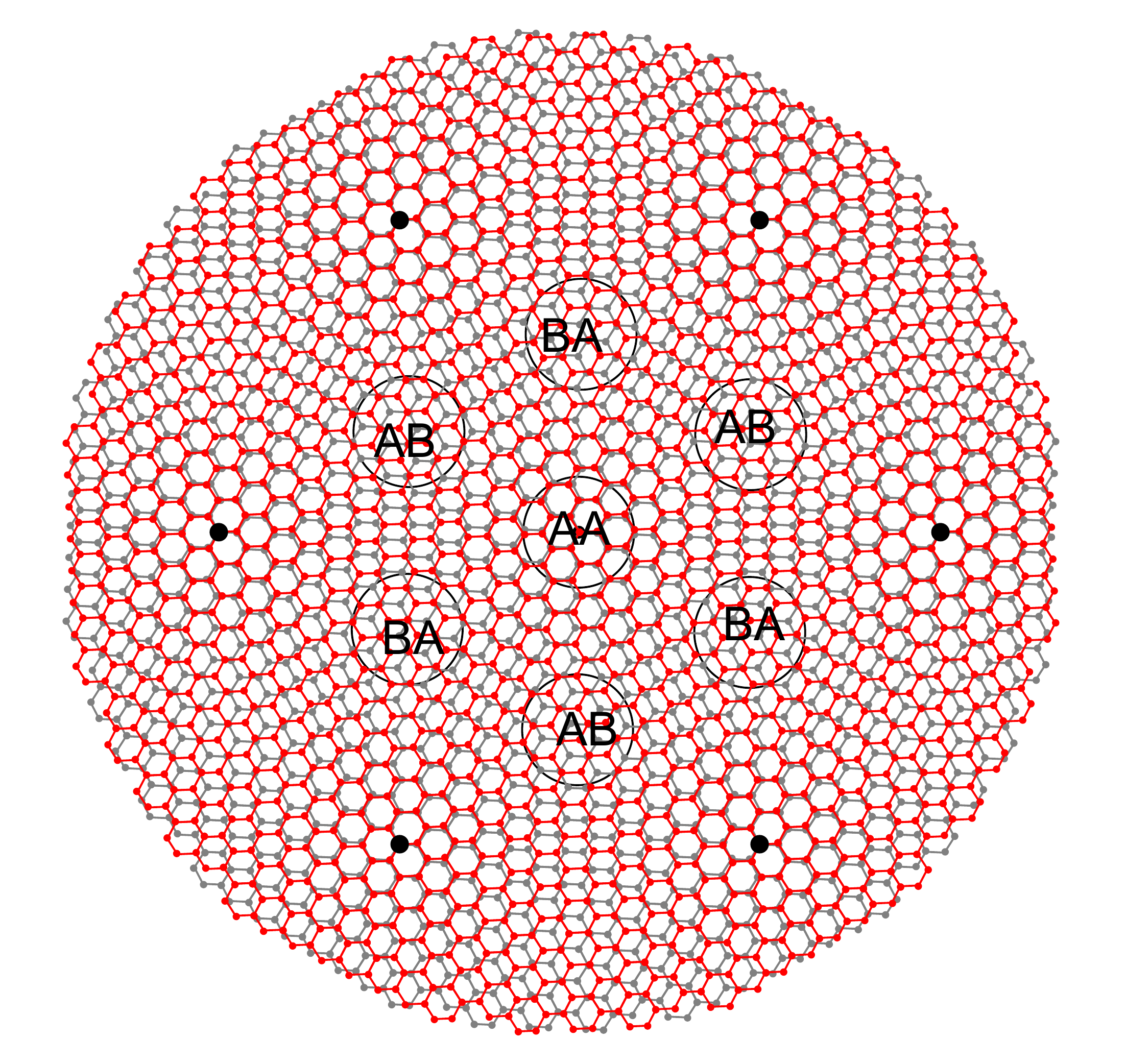}
\caption{Illustration of twisted bilayer graphene system in real space. The small twist angle $\theta$ has a commensurate value $\theta = \arccos\frac{253}{254} \simeq 0.0888$, with $(m,n) = (6,7)$ (see main text for the definition of comensurate angle, $m$ and $n$). Red (gray) is for the top (bottom) layer. The origin is chosen at an $AA$ region center (perfect $AA$ stacking point). The $AB$ and $BA$ regions closest to the origin are labeled. Further $AA$ stacking points are shown in black dots.}\label{illus}
\end{figure}

A single-layer graphene is a 2D honeycomb lattice of carbon atoms. Its periodic unit belongs to the triangular type, with a unit cell consisting of two atoms which we call $A$ and $B$. The twisted bilayer graphene (TBG) is one layer of graphene sitting on top of the other, whose difference in orientations is parameterized by an angle $\theta$. When $\theta$ is small, the TBG system forms long interference pattern which is called moir\'e patterns. While there is no unit cells for a generic $\theta$, the TBG exhibits an $m\times n$ enlarged unit cell at specific value of $\theta$ given by $\cos \theta = \frac{m^2+4mn+m^2}{2(m^2+mn+n^2)}$ \cite{2007PhRvL..99y6802L}, where $m,n$ are arbitrary positive integers, and the corresponding $\theta$ is called a commensurate angle (see Fig. \ref{illus}). 

The continuum model~\eqref{eq:TBGa} introduced in the main text is a low energy model applicable to both commensurate and incommensurate cases. 
A more general form of this model writes
\begin{eqnarray}
\label{eq:TBGb}
H_\xi(\bm{x}) = &&-i\,\mathrm{R}\left(\frac{\tau^z \theta}{2}\right)\left(\bm{\nabla}  + i\xi \left(\bm{q}_{\text{h}}-\tau^z \frac{\bm{q}_0}{2}\right)\right)\cdot(\xi\sigma^x,\sigma^y)
\nonumber\\
&&+ \alpha \; \tau^+ \left[\alpha(\bm{x})+\beta(\bm{x}) \sigma^+ + \gamma(\bm{x}) \sigma^-\right] +\mathrm{h.c.},\nonumber\\
\end{eqnarray}
where $\xi = \pm 1$ labels the two inequivalent valleys, and $\mathrm{R}(\theta)$ is the standard 2D rotation matrix that rotates the momenta by $\theta$ in the plane, in response to the real space twist between graphene layers. Note that one needs to use $\xi \bm{Q}_i$ and $\zeta^\xi$ instead of $\bm{Q}_i$ and $\zeta$ in the definitions of $\alpha(\bm{x})$, etc. At small angles in TBG the physics near the $\xi=\pm1$ valleys are decoupled from each other, however, either a $C_2$ rotation or a time reversal transformation $\mathcal{T}$ maps one into the other. Furthermore note that the rotation of the momenta introduced above by the rotation mtrix $\mathrm{R}$ is equivalent to a rotation of the sublattice sigma matrices instead; in the main text we have addressed this point by the rotation of the sublattice matrices in the Dirac term. The main text has chosen to focus only on $\xi=1$ and neglected the small $\theta$ rotation for the momentum.

\section{Topological charge of singular point}
\label{sec:berry_phase}
In this Appendix, we describe the numerical method to compute the vorticity (topological charge) of a singular point in the band structure, such as a Dirac point. We follow the method introduced in Ref.~\onlinecite{Fukui2005}.

 The topological charge around a singular point is defined as a contour integral around this point
\begin{align}
N=\frac{1}{2\pi}\oint_{\mathcal{C}}\vec{\mathcal{A}}\cdot d\vec{l}=\frac{1}{2\pi}\int_{\Sigma}\vec{\mathcal{B}}\cdot\hat{n} dS,
\end{align}
where the Berry connection $\mathcal{A}_j$ is given in terms of a derivative 
\begin{align}
\mathcal{A}_j=-i \langle \psi({\bm k})|\partial_j|\psi({\bm k})\rangle
\end{align}
and Berry curvature is defined in terms of a second derivative, i.e., $\mathcal{B}=\partial_i\mathcal{A}_j-\partial_j\mathcal{A}_i$.
For DP, the topological charge can take the values $N=\pm 1/2$. The Berry curvature is zero except at the DP, where $\mathcal{B}$ diverges to either $\infty$ or $-\infty$.

To resolve the divergence of Berry curvature at band crossing point, we use the lattice Berry curvature introduced in Ref.~\onlinecite{Fukui2005}. In the case of a singly occupied band, the single-particle Berry curvature is given by
\begin{align} \label{eq:chern}
\mathcal{B}_\ell &=i \log U_1({\bm k}_\ell) U_2({\bm k}_\ell+\delta_x)  U_1({\bm k}_\ell+\delta_y)^{-1} U_2({\bm k}_\ell)^{-1},
\end{align}
where ${\bm k}=(k_x,k_y)$ is defined in the first Brillouin zone. $U_{\mu}$ is the Berry connection and is given by 
\begin{align} \label{eq:berry_band}
U_\mu({\bm k}_\ell)= \frac{\braket{u ({\bm k}_\ell) | u ({\bm k}_\ell+\delta_\mu) }}
{\left| \braket{u ({\bm k}_\ell) | u ({\bm k}_\ell+\delta_\mu) } \right|},
\end{align}
where $\ket{u({\bm k}_\ell)}$ is the Bloch function of the occupied state, i.e. eigenstate of the Hamiltonian in momentum space $h({\bm k}) \ket{u({\bm k}_\ell)}= \epsilon_{\bm k} \ket{u({\bm k}_\ell)}$.

If there are $N$ occupied bands, then the connection is given by
\begin{align} \label{eq:berry_bands}
U_\mu({\bm k}_\ell)= \frac{\text{Det}_{mn}\left[ {\langle u_m({{\bm k}_\ell}) |u_n({{\bm k}_\ell+\delta_\mu})\rangle}\right]}
{\left|\text{Det}_{mn}\left[ {\langle u_m ({{\bm k}_\ell}) | u_n ({{\bm k}_\ell+\delta_\mu}) \rangle}\right]\right|},
\end{align}
where $\ket{u_n({\bm k}_\ell)}$ is the $n$-th occupied state at momentum ${\bm k}_\ell$. A proof of the above formula follows from the anti-commutation relation of fermion operators and the Wick's theorem.

In this paper, we use the above formula to study the topological charge of band crossing points in the flat band in the vicinity of the first magic angle.

\subsection*{Moir\'e lattice}

Here we first show that in the Bistritzer-Macdonald's model\cite{Bistritzer2011}, $C_2 {\cal T}$ symmetry forces the Berry curvature to be zero except at band crossing points.
Let us begin with the definition of the Berry curvature at point ${\bm k}$,
\begin{align}
\mathcal{B}({\bm k}) =i\epsilon_{ij}  \partial_i \braket{ u({\bm k})| \partial_j| u({\bm k})},
\end{align}
where $\partial_j :=\partial /\partial k_j$.

We now look at the Berry curvature of the transformed state
\begin{align}
\ket{\tilde{u}({\bm k})}=C_2 {\cal T} \ket{u({\bm k})}= U_{C_2 {\cal T}} \ket{u({\bm k})}^\ast
\end{align}
where $U_{C_2 {\cal T}}=\sigma^x$ is the unitary part of the transformation and the asterisk means complex conjugation. Note that $C_2 {\cal T}$ sends ${\bm k} \to {\bm k}$, unlike pure ${\cal T}$.
We have
\begin{align}
\mathcal{B}({\bm k}) & = i\epsilon_{ij} \partial_i \braket{ u({\bm k})| (C_2 {\cal T})^{-1} (C_2 {\cal T}) \partial_j  |u({\bm k})} \nonumber \\
&= i\epsilon_{ij} \partial_i \ce{{^\ast}\braket{  u({\bm k}) | U^\dag_{C_2 {\cal T}} \partial_j U_{C_2 {\cal T}} | u({\bm k})}^\ast} \nonumber  \\
&= i\epsilon_{ij} \partial_i \braket{  u({\bm k}) |  \overleftarrow{\partial_j}  | u({\bm k})} \nonumber  \\
&= - i\epsilon_{ij} \partial_i \braket{  u({\bm k}) |  \partial_j  | u({\bm k})} \nonumber  \\
&= - \mathcal{B}({\bm k}) 
\end{align}
which implies $\mathcal{B}({\bm k}) =0$, unless there is a band crossing.

\begin{figure}
\centering
\includegraphics[width=.4\textwidth]{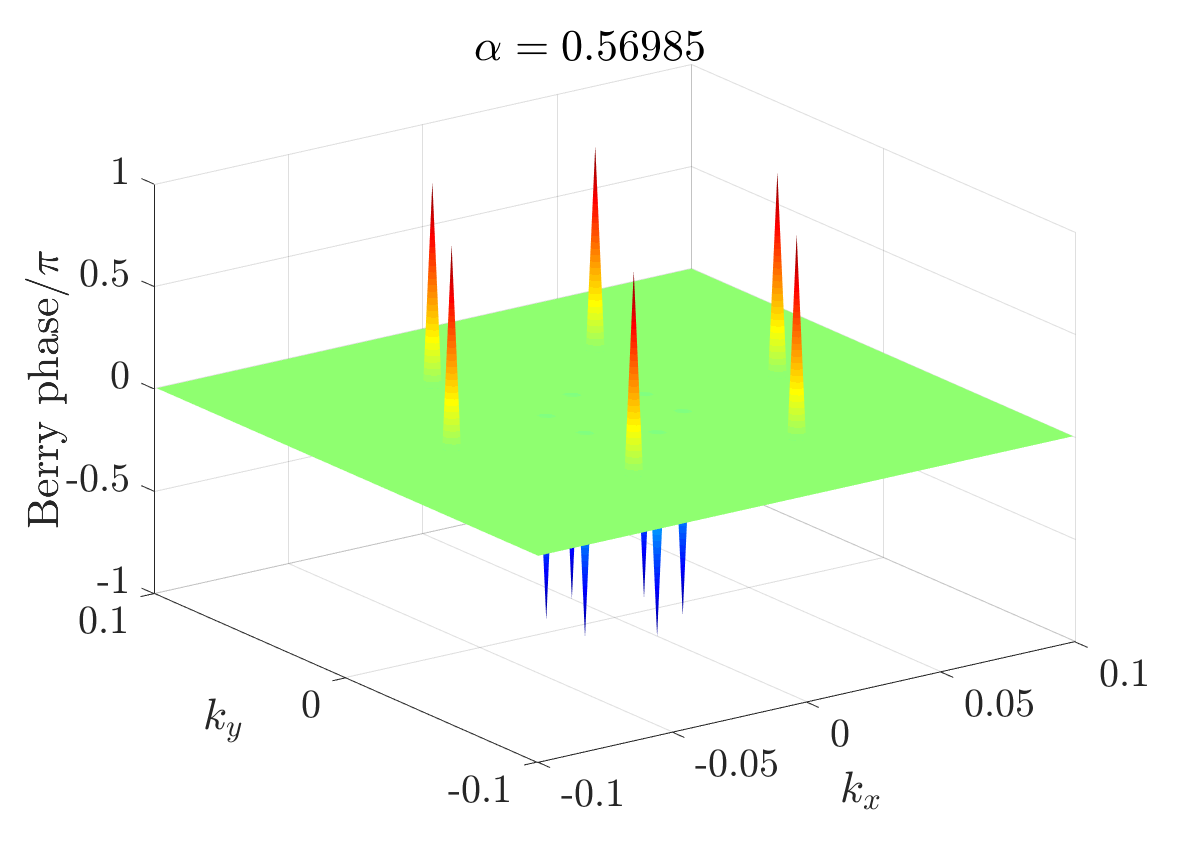}
\caption{The Berry phase of the 12 DPs around $\Gamma$ point at $\alpha=0.56985$. The Berry phase is either $\pi$ or $-\pi$.}
\label{fig:berry_alpha_056985}
\end{figure}

As mentioned, $\mathcal{B}({\bm k})$ diverges at band crossing points. Using Eq.~(\ref{eq:berry_band}), we actually  evaluate the Berry phase around a singular point. Traversing a path around a DP leads to a Berry phase of $\pm \pi$ (See Fig. \ref{fig:berry_alpha_056985}). 
In order to resolve the $2\pi$ ambiguity of the lattice Berry curvature using Eq.~(\ref{eq:berry_band}), we added an infinitesimal $C_2 {\cal T}$ breaking term, $\delta \,  \sigma^z$, to the CM Hamiltonian where $\delta \sim 10^{-10}$. This small term acts as a reference mass term near each DP and resolves the Berry phase ambiguity of the massless DPs. Note that the resulting charge of each DP does depend upon the form of the global mass term, but we take to be true that the relative charges do not depend on the form of the global mass term.

\section{The effective six-band model}
\subsection{Schrieffer-Wolff transformation}\label{sec:SW_two_band}
In this section of this appendix, we outline how the effective two-band model shown in Eq. \eqref{eq:eff_two_band} is derived from the full six-band Hamiltonian \eqref{eq:six_band_hamiltonian_matrix} using the Schrieffer-Wolff (SW) transformation. We will use the notation and approach used in Appendix B of Ref.~\onlinecite{slagle2017fracton}.

 Our goal is to carry out the SW effective Hamiltonian in a perturbative (in powers of quasi-momentum) fashion; we take the unperturbed Hamiltonian to have the following form:
 \begin{align}
H_6^{(0)}=\begin{pmatrix} \Delta \, \tau^z + v \, \bm{k}\cdot \bm{\sigma} & 0\\
0 & \delta \, \mu^z\end{pmatrix},
\end{align}
while treating what remains as a perturbation:
\begin{align}
H_6^{(1)}= \lambda \begin{pmatrix} d_2\left(k^2_+\sigma^+ + k^2_- \sigma^-\right) \tau^z & H_{42}\\
H_{24} & b_1 k^2 \, \mu^z\end{pmatrix}.
\end{align}
 Note that the matrix $H_6^{(0)}$ is a six-band Hamiltonian consisting of decoupled two-dimensional and four-dimensional sectors; in $H_6^{(1)}$, on the other hand, we have introduced the factor $\lambda$ to keep track of the powers of perturbation and we have neglected $d_1$ and $d_3$ as explained in the next section of this appendix. Since we are interested in the regime of very small $\delta$, we will perform the perturbative calculation assuming $\delta=0$; this means that we are left with a degenerate low-energy subspace and we need to utilize degenerate perturbation theory. Neglecting it in the perturbative calculations, we will add the $\delta \mu^z$ term to the final Hamiltonian eventually.

We seek an anti-Hermitian operator $S$ which can be exploited to define an effective Hamiltonian as follows,
\begin{equation}
	H_6^{\text{eff}} = e^S \, H \, e^{-S},
\end{equation}
such that 
\begin{equation}\label{eq:cummutator_relation}
	\left[ H_6^{\text{eff}} , \mathcal{P} \right] = 0,
\end{equation}
where $\mathcal{P}$ is the projector onto the two-dimensional low energy subspace. The commutation relation above, when satisfied, means that $H_6^{\text{eff}}$ consists of two decoupled low energy and high energy sectors. We find $S$ and $H_6^{\text{eff}}$ as power series in the perturbation parameter $\lambda$, so that the commutation relation \eqref{eq:cummutator_relation} is satisfied at each order, according to a prescription given in Ref.~\onlinecite{slagle2017fracton}.

$H_6^{\text{eff}}$ is found to fifth order in $\lambda$, this guarantees that the result is accurate to sixth order in $k$; however, it is not in the form of a power series in $k$, since $H_6^{(0)}$ also contains $k$; thus the result is expanded to sixth order in $k$. The effective two-band  Hamiltonian will then follow:
\begin{equation}
\begin{aligned}
	H_2^{\text{eff}}  &= \mathcal{P} H_6^{\text{eff}} \mathcal{P}  \\
	&= c_0 \; \mathrm{Re}(k^3_+) \\
	&+ \left( \delta + c_{3,2} \; k^2 + c_{3,6} \; \mathrm{Re}(k^6_+) \right) \mu^z \\
	&+ c_{1,6} \; \mathrm{Im}(k^6_+)\;\mu^x,
\end{aligned}
\end{equation}
where only the lowest order correction to each angular-pseudospin dependence is retained here. The coefficients can be given in terms of the original \textit{bare} parameters as:
\begin{equation}
\begin{aligned}
	c_0 &= \frac{2 \left(g_1^2-g_2^2\right) v}{\Delta ^2} ,\\
	c_{3,2} &= - \frac{2 \left(g_1^2-g_2^2\right) }{\Delta },\\
	c_{3,6} & = \frac{2 \left(g_1^2+g_2^2\right) v^2 \left(d_2 \Delta + 2 g_1^2-2 g_2^2\right)}{\Delta ^5} , \\
	c_{1,6} & = -\frac{4 g_1 g_2 v^2 \left(d_2 \Delta +2 g_1^2-2 g_2^2\right)}{\Delta ^5}.
\end{aligned}
\end{equation}
Their numerical values can be derived according to the values shown in Eq.~\eqref{eq:parameters_2nd_appendix} in Appendix \ref{sec:parameter_six_band}:
\begin{equation}
\begin{aligned}
	c_0 &= 8.4,  \\
	c_{3,2} &= -0.24, \\
	c_{3,6} &= 2.3 \times 10^6, \\
	c_{1,6} &= -2.3 \times 10^6 .
\end{aligned}
\end{equation}
Note that we will be dealing with very small values of $k$ ($k \ll 0.01$) in this effective Hamiltonian and that is why some coefficients turn out to be large.

\subsection{Parameters in the six-band model}\label{sec:parameter_six_band}
In this section of this appendix, we discuss how the parameters in the effective Hamiltonian are derived. Let us inform the reader that we have not attempted to derive error bars for the following parameter values. 

First, in the effective model with linear terms only, the dispersion relations to second order in $k$ can be written as:
\begin{equation}
	\begin{aligned}
		\epsilon_{0,\pm} &= \pm \left[\delta + 2 k^2 \left(\frac{g_1^2}{\delta -\Delta }+\frac{g_2^2}{\delta +\Delta }\right) \right],\\
		\epsilon_{-1,\pm} &= -\Delta \pm k v \\
		& \qquad + k^2 \left(\frac{g_1^2 \; (1 \pm \cos 3 \theta_{\bm{k}} )}{\delta -\Delta }-\frac{g_2^2 \; (1 \mp \cos 3 \theta_{\bm{k}} )}{\delta +\Delta }\right),\\
		\epsilon_{+1,\pm} &= \Delta \pm k v \\
		& \qquad + k^2 \left(-\frac{g_1^2 \; (1 \pm \cos 3 \theta_{\bm{k}})}{\delta -\Delta }+\frac{g_2^2 \; (1 \mp \cos 3 \theta_{\bm{k}} )}{\delta +\Delta }\right), 
	\end{aligned}
\end{equation}
where $\theta_{\bm{k}} = \arctan\left( k_y / k_x\right)$ is the angle measured from the $k_x$ direction. The indices are motivated by the spacing of the above energy levels at $\alpha = \alpha_1$.

 In the main text, we have shown that $\Delta=S(\alpha_2-\alpha)$ and $\delta=s(\alpha-\alpha_1)$,
with $s = 1.13$ and $S = 1.29$. To find the rest of the parameters, we focus on $\alpha = \alpha_1$ and extract these parameters from the properties of the numerical dispersion curves at this point. We list the dominant $\boldsymbol{k}$ dependent behaviors in certain combinations below:
\begin{itemize}
\item The sum of two terms such as $\epsilon_{-1,+}$ and $\epsilon_{+1,+}$ when $\boldsymbol{k}$ is in the $k_y$ direction is simply equal to $2 v k$.
\item The difference between $\epsilon_{0,+}$ and $\epsilon_{0,-}$ is given by $4 k^2\left( g_1^2 - g_2^2 \right)/\Delta$.
\item The difference between two terms such as $\epsilon_{-1,+}$ and $\epsilon_{+1,+}$ along $k_x$ direction (with fixed $k_y=0$) is $2\left( \Delta + 2 k^2 g_1^2/\Delta \right)$.
\end{itemize}

By fitting to graphs of the above combinations in well-chosen intervals, one is able to find the following values of linear parameters:
\begin{equation}
	v = 0.263, \qquad g_2 = 0.339, \qquad g_1 - g_2 = 0.00146.
\end{equation}

Second, we show how the coefficients of the quadratic model can be determined by fitting to suitable combinations of energy functions. The quadratic coefficients include $b_1,d_1,d_2,d_3$. With the introduction of quadratic corrections, the very small $k$ expansions of the six energies become:
\begin{equation}
	\begin{aligned}
		\epsilon_{0,\pm} &= \pm  \left[ \delta + 2 k^2 \left(b_1 + \frac{g_1^2}{\delta -\Delta }+\frac{g_2^2}{\delta +\Delta }\right) \right],\\
		\epsilon_{-1,\pm} &= -\Delta \pm k v + k^2 \bigg(- d_1 \mp d_2 \cos 3\theta_{\bm{k}}  \\
		& \qquad + \frac{g_1^2 \; (1 \pm \cos 3 \theta_{\bm{k}} )}{\delta -\Delta }-\frac{g_2^2 \; (1 \mp \cos 3 \theta_{\bm{k}} )}{\delta +\Delta }\bigg),\\
		\epsilon_{+1,\pm} &= \Delta \pm k v + k^2 \bigg( d_1 \pm d_2 \cos 3\theta_{\bm{k}}  \\
		& \qquad -\frac{g_1^2 \; (1 \pm \cos 3 \theta_{\bm{k}})}{\delta -\Delta }+\frac{g_2^2 \; (1 \mp \cos 3 \theta_{\bm{k}} )}{\delta +\Delta }\bigg). 
	\end{aligned}
\end{equation}

Similar to the linear case, we will focus on $\alpha = \alpha_1$ to determine the coefficients. As is stated in the main text, the difference between the two middle bands, shows a further quadratic behavior in the regime of larger $k$, which is not captured by the linear effective six-band model, the linear model shows saturation in that regime. We would like to use some of the information given by this quadratic behavior along with the above very small $k$ dispersion relations;
although combinations of the above energies give enough equations for all the quadratic coefficients, one should be careful, since in the above forms, the quadratic coefficients are added to terms that are orders of magnitude larger. 

To understand this saturation behavior in the linear effective model, consider $\alpha = \alpha_1$ for concreteness; the regime $|\bm{k}| \gtrapprox 0.02$, i.e.~saturation in $\Delta E$ of the two middle bands, corresponds to the situation in which terms $v k$ and $g_i k$ are not perturbations to $\Delta$ anymore; if one goes to higher and higher values of $k$, the latter actually becomes a perturbation to the linear terms and $\frac{\Delta}{vk}$ turns out to be a small parameter. Indeed, it is not correct to think of the effective system as composed of the two-band and the four-band sectors, being weakly coupled in this regime; all the eigenvectors are in fact non-perturbative superpositions of the two-band and four-band basis vectors.

Motivated by the above discussion, we seek to find the behavior for the regime $k v \gg \Delta$ for all of the six bands. To this end, we impose $\Delta \to 0$ and $ g_1 \to g_2$ and find the dominant $k$ dependent terms for the six bands treating the quadratic correction as a perturbation to the linear terms:
\begin{eqnarray}
		&&\epsilon_{0,\pm} =  \pm k^2  \times\nonumber\\
		&&\frac{\sqrt{b_1^2 v^4 + 8 b_1 g_2^2 v^2 (d_2 + d_3) \cos 6\theta_{\bm{k}} + 16 g_2^4 (d_2 + d_3)^2}}{4 g_2^2 + v^2},\nonumber\\
		&&\epsilon_{-1,\pm} =  -\left| k \right| \sqrt{4g_2^2 + v^2},\nonumber\\
		&&\epsilon_{+1,\pm} =  \left| k \right| \sqrt{4g_2^2 + v^2}.
\end{eqnarray}
Note that in the expressions for $\epsilon_{0,\pm}$ only the combination $d_2 + d_3$ appears; also, having the fact that the quantities $d_2 - d_3$ and $d_1$ only appear in the subdominant quadratic coefficients of $\epsilon_{\pm 1}$ (which are not shown here) in mind, we impose $d_1 = d_3 = 0.$
The following dominant $k$ dependent behaviors at $\alpha = \alpha_1$ can be used for finding the coefficients including the quadratic ones:
\begin{itemize}
	\item The sum of two terms such as $\epsilon_{-1,+}$ and $\epsilon_{+1,+}$ for very small $k$, i.e.~$k v \ll \Delta$, when $\boldsymbol{k}$ is in the $k_y$ direction is again equal to $2 v k$.
	\item The difference between $\epsilon_{0,+}$ and $\epsilon_{0,-}$, for very small $k$, i.e.~$k v \ll \Delta$ is now given by $4 k^2 \left[ - b_1 + \frac{1}{\Delta} \left( g_1^2 - g_2^2 \right)\right]$.
	\item The difference between two terms such as $\epsilon_6$ and $\epsilon_4$ for very small $k$, i.e.~$k v \ll \Delta$, when $\boldsymbol{k}$ is in the $k_x$ direction now reads $2\left[ \Delta +  k^2 \left(d_2 + 2\frac{g_1^2}{\Delta} \right)\right]$.
	\item The coefficient of the quadratic term of the difference between $\epsilon_{0,+}$ and $\epsilon_{0,-}$ for the regime $k v \gg \Delta$, has the form 
	$$\frac{2\sqrt{b_1^2 v^4 \pm 8 b_1 g_2^2 v^2 (d_2 + d_3) + 16 g_2^4 (d_2 + d_3)^2}} {\left(4 g_2^2 + v^2\right)},$$ 
	where the $+(-)$ sign is chosen when $\boldsymbol{k}$ is in the $x(y)$ direction.
\end{itemize}
Again by fitting to graphs of the above combinations in well-chosen intervals, one is able to find the following values for the parameters:
\begin{eqnarray}\label{eq:parameters_2nd_appendix}
	&&v = 0.263, \qquad g_2 = 0.339, \qquad g_1 - g_2 = 0.00130, \nonumber\\
	&&\qquad b_1 = -0.0289, \qquad d_2 = -0.0106.
\end{eqnarray}
Note that with the introduction of quadratic terms, the previously determined coefficients have also changed.

\bibliography{graphene}

\end{document}